\documentclass[longauth]{aa}  
\usepackage[utf8]{inputenc}
\usepackage{comment}
\usepackage{enumitem}
\usepackage{graphicx}	
\usepackage{caption}
\usepackage{subcaption}
\usepackage{xcolor}
\usepackage{gensymb, amsmath,amssymb}
\usepackage{booktabs}
\usepackage[hidelinks,colorlinks=true,linkcolor=blue,citecolor=blue,urlcolor=black]{hyperref}
\usepackage{lscape}
\usepackage{makecell}
\begin{document}

    \title{CHEX-MATE: Probing the thermal–non-thermal connection in radio halo clusters with LoTSS-DR2}
    
    
    \titlerunning{CHEX-MATE: X-ray -- radio study on LoTSS DR2 clusters}
    \authorrunning{M. Balboni et al.}
    
    \author{
    M. Balboni \inst{\ref{unibo}, \ref{iasf},}
    \and F. Gastaldello \inst{\ref{iasf}}
    \and A. Bonafede \inst{\ref{unibo},\ref{ira}}
    \and Y. E. Bahar \inst{\ref{inaf-oas}}
    \and I. Bartalucci\inst{\ref{iasf},\ref{UniR2}}
    \and A. Botteon\inst{\ref{ira}}
    \and H. Bourdin\inst{\ref{UniR2},\ref{INFN-R2}}
    \and G. Brunetti\inst{\ref{ira}}
    \and R. Cassano\inst{\ref{ira}}
    \and F. De Luca\inst{\ref{UniR2},\ref{INFN-R2}}
    \and S. Ettori\inst{\ref{inaf-oas},\ref{infn-bo}}
    \and M. Gaspari\inst{\ref{unimore}}
    \and M. Gitti\inst{\ref{unibo},\ref{ira}}
    \and A. Ignesti\inst{\ref{cze_sa}}
    \and C. Jones\inst{\ref{cfa}}
    \and L. Lovisari\inst{\ref{iasf},\ref{cfa}}    
    \and P. Mazzotta\inst{\ref{UniR2},\ref{INFN-R2}}
    \and E. Pointecouteau\inst{\ref{irap}}
    \and G. W. Pratt\inst{\ref{saclay}}
    \and M. Rossetti\inst{\ref{iasf}}
    \and J. Sayers\inst{\ref{caltec}}
    \and M. Sereno\inst{\ref{inaf-oas},\ref{infn-bo}}
    \and R. J. van Weeren\inst{\ref{leiden}}
    \and S. De Grandi\inst{\ref{brera}}
    \and S. Ghizzardi\inst{\ref{iasf}}
    }
    %
    \institute{
        DIFA - Universit\`a di Bologna, via Gobetti 93/2, I-40129 Bologna, Italy\label{unibo}
        \and
        INAF - IASF Milano, via A. Corti 12, 20133 Milano, Italy \label{iasf}
        \and
        INAF - IRA, Via Gobetti 101, I-40129 Bologna, Italy\label{ira}
        \and
        INAF - Osservatorio di Astrofisica e Scienza dello Spazio di Bologna, via Piero Gobetti 93/3, I-40129 Bologna, Italy\label{inaf-oas}
        \and
        INFN, Sezione di Bologna, viale Berti Pichat 6/2, I-40127 Bologna, Italy \label{infn-bo}
        \and
        California Institute of Technology, Pasadena, CA 91125, USA\label{caltec}
        \and
        Dipartimento di Fisica, Università di Roma ‘Tor Vergata’, Via della Ricerca Scientifica 1, I-00133 Roma, Italy.\label{UniR2}
        \and
        INFN, Sezione di Roma ‘Tor Vergata’, Via della Ricerca Scientifica, 1, 00133, Roma, Italy.\label{INFN-R2}
        \and
        Department of Physics, Informatics and Mathematics, University of Modena and Reggio Emilia, 41125 Modena, Italy\label{unimore}
        \and
        Universit\'e Paris-Saclay, Universit\'e Paris Cit\'e, CEA, CNRS, AIM, 91191 Gif-sur-Yvette, France\label{saclay}
        \and
        IRAP, CNRS, Universit\'e de Toulouse, CNES, Toulouse, France\label{irap}
        \and
        Center for Astrophysics $\rvert$ Harvard \& Smithsonian, 60 Garden Street, Cambridge, MA 02138, USA\label{cfa}
        \and
        Astronomical Institute of the Czech Academy of Sciences, Bo\v cn\'i II 1401, 14100 Prague, Czech Republic\label{cze_sa}
        \and
        Leiden Observatory, Leiden University, PO Box 9513, 2300 RA Leiden, The Netherlands\label{leiden}
        \and
        INAF - Osservatorio Astronomico di Brera, via E. Bianchi 46, I-23807 Merate (LC), Italy\label{brera}
        }

   \abstract{
   The main baryonic component of galaxy clusters is in the form of a hot plasma observed in the X-rays. In addition, clusters host a non-thermal component (relativistic particles and magnetic fields) which is studied mainly via radio telescopes and is dynamically coupled to the thermal plasma.
   These two components evolve together within the system and are subject to both micro and macro-physical processes which can be traced only with deep and detailed observations.
   We present the first homogeneous spatially resolved radio--X-ray study of 18 radio halo clusters combining CHEX-MATE and the LOFAR Two metre Sky Survey DR2.
   Radio ($I_R$) and X-ray ($I_X$) surface brightness are clearly correlated (Spearman coefficient $r_s \gtrsim 0.5$), following a typically sublinear $I_R$–$I_X$ relation ($67 \pm 5\%$ of the sample) that appears to be independent of other cluster properties, although a tentative trend with mass/redshift may be worth exploring with larger samples.
   When inspecting such a relation as a function of the radius, we observe a variety of behaviours, which suggest that the interplay between the thermal and non-thermal plasma components varies with radius.
   We find the averaged slope profile of dynamically disturbed clusters to be consistent with expectations from turbulent re-acceleration models where a constant ratio between cosmic-ray electron and thermal energy densities is assumed.
   Differently, less perturbed clusters require an increasing cosmic-ray-to-thermal energy ratio with radius.
   Finally, we show that radio halo surface-brightness profiles can be effectively rescaled using X-ray information and theoretical expectations. This procedure significantly reduces their scatter and reveals common underlying trends based on the cluster dynamics, resembling what is observed for the thermal ICM, and strengthening the idea of a strong interplay between these two plasma components.
   }
   
\maketitle
\section{Introduction}\label{sec:intro}

Galaxy clusters allow us to study a wide range of processes, from structure formation and cosmology to particle re-acceleration processes and radiative mechanisms.
These two latter subjects are mainly explored through radio and X-ray observations. In fact, it is well established that within the cluster environment, many sources of radio synchrotron emission are present \citep[see e.g.][for a review]{vanWeeren19}. 
Among the most puzzling and extended cases of such emission are giant radio halos.
They are megaparsec-scale sources characterised by a steep spectral index ($\alpha < -1$, where the flux density $S_{\nu} \propto \nu^{\alpha}$) mainly found in massive merging clusters.
Although two main mechanisms for the production of radio halos have been proposed (hadronic and leptonic origin), past gamma ray observations have discarded a pure hadronic origin of the cosmic ray electrons (CRe, see \citealt{Brunetti14} for a review).
Instead, statistical studies on radio halos at low frequencies showed an agreement with the expectations from the (re-)acceleration of CRe operated by merger-induced turbulence \citep[e.g.][]{Cassano2023}.
In fact, during the structure formation process, galaxy clusters experience the accretion of sub-clusters and groups through mergers, which release up to $\sim 10^{64}$ ergs in the Intra Cluster Medium (ICM) in a cluster crossing time \citep[$\sim$ Gyr, e.g.][]{Tormen2004}. 
Such energy is mainly dissipated into gas heating through shocks, enhancing the ICM thermal Bremsstrahlung emission in the X-rays. A small fraction of this energy can be channelled into turbulence that re-accelerates relativistic electrons via a Fermi-II like mechanism, originating synchrotron emission \citep{Brunetti2001,Petrosian2001}. 
Since cluster mergers are ultimately driven by gravity, the energy available for particle acceleration is expected to scale with the cluster gravitational potential, and hence with its mass. Consequently, the most massive mergers are expected to generate bright radio halos observable up to GHz frequencies, whereas less energetic mergers should preferentially produce fainter halos with steeper spectra, detectable mainly at low radio frequencies.
These latter objects have indeed been studied in both targeted observations \citep[e.g.][]{Macario13, Wilber2018} and large samples \citep{Cassano2023}.\\
This turbulent mechanism for halo production reproduces well the general properties of radio halos, like their number and redshift distribution or the expected flux \citep{Cassano2023}. It also provides a natural explanation for the observed radio power - mass relation and the halo merger connection \citep[e.g.][]{Cassano2010b,Cuciti2023,DiGennaro2025}. However, the spatially resolved properties of such a process remain poorly understood.
In fact, different turbulence re-acceleration mechanisms (transit time damping;  Alfven resonant scattering, etc.) predict different spatial distributions of relativistic particles, magnetic fields, and thermal plasma, leading to a variety of observational signatures and physical scenarios \citep[e.g.][]{Brunetti-Lazarian07, Brunetti-Lazarian11, Miniati15, Brunetti-Lazarian16}.
Thanks to the high sensitivity of current radio facilities (e.g., LOFAR, \citealt{LOFAR2013}, MeerKAT, \citealt{Jonas2016}, ASKAP, \citealt{Hotan2021}), spatially resolved studies of radio and X-ray derived quantities are becoming possible for large samples of objects. They have the great potential to infer the details of the radio emission mechanism (e.g. CRe distribution, acceleration rate, magnetic field) and to put constraints on the microphysics of the ICM \citep[e.g.][]{Govoni2001, Kamlesh2021-A2744, Balboni2024}.\\
Recent works exploited the correlation among radio and X-ray surface brightness in the centre and in the periphery of clusters to derive constraints on the non-trivial distribution of the synchrotron emitting electrons \citep[e.g.][]{Biava21, Biava2024}. Similarly, from radio spectral index profiles it is now possible to constrain some of the crucial parameters of the model, such as the typical re-acceleration time \citep[$t_{\text{acc}}$, e.g.][]{Kamlesh2023-A2256, Kale2025}.\\ 
Studies on large samples of halos have shown that their global properties are broadly consistent with the predictions of homogeneous re-acceleration models, based on cluster-averaged physical quantities \citep{Cassano2023}.
However, when radio spectral index profiles are available and allow to constrain some of the crucial parameters of the model such as $t_{acc}$, radial variations emerge, indicating that these averaged descriptions may not fully capture the underlying physics \citep[e.g.][]{Kamlesh2023-A2256,Bonafede2022}.\\
Only a handful of detailed studies combining spatially resolved radio and X-ray observations have so far explored these effects.
\cite{Bonafede2022} (and subsequently \citealt{Bruno2023} and \citealt{Balboni2024}) investigated the radial trend of the radio and X-ray surface brightness relation, finding evidence that the ratio between thermal and non-thermal energy densities may vary with radius.
Similarly, \cite{Santra2024} and \cite{vanWeeren2026} also provided evidence for a non-trivial distribution of CRe throughout the cluster extension, and for different CRe production scenarios. 
These studies demonstrate the potential of spatially resolved analyses, but remain limited to a small number of individual clusters. A homogeneous analysis of a statistically representative sample is therefore still lacking.\\
With the present work, we aim to perform the first systematic study of radio halo resolved brightness properties and their correlations with thermal ICM emission.
In particular, we aim to study a well selected sample of objects, for which we have uniform and homogeneous coverage in both radio and X-ray bands. 
This is possible by matching the Cluster HEritage project with XMM-Newton - Mass Assembly and Thermodynamics at the Endpoint of structure formation \citep[CHEX-MATE,][]{Arnaud21} and the LOFAR Two-meter Sky Survey Data Release 2 \citep[LoTSS-DR2,][]{Shimwell2022} datasets. They provide the necessary data in the two requested bands and for a wide sample of Planck 2nd Sunyaev-Zeldovich Source Catalog objects \citep{Planck2016}.\\
The paper is organized as follows: in Section~\ref{sec:obs-data} we present the sample observations and data reduction; 
in Section~\ref{sec:data_analysis} we describe the data analysis; in Section~\ref{sec:discussion} we comment on our results. In Sections~\ref{sec:summary} we summarise the results.\\
In this work we assume a flat, $\Lambda$CDM Universe cosmology with $H_0 = 70 \rm ~ km ~ s^{-1} ~ Mpc^{-1}$, $\Omega_{m,0}=0.3$, $\Omega_{\Lambda} = 0.7$, $H_z / H_0 = E_z = [ \Omega_{m, 0} (1 + z)^3 + \Omega_{\Lambda} ]^{1/2}$.
\section{Sample observations and data reduction}\label{sec:obs-data}
\subsection{CHEX-MATE}

The CHEX-MATE project\footnote{\protect\url{http://xmm-heritage.oas.inaf.it/}} \citep{Arnaud21} collected three mega-seconds of XMM-Newton observations for a minimally-biased, signal-to-noise limited sample of 118 galaxy clusters detected by Planck through the Sunyaev-Zel'dovic effect \cite{SZ1972}.
The program aims to study the ultimate products of structure formation in time and mass by selecting the most recent objects to have formed (Tier-1: $0.05 < \rm{z} < 0.2$ ; $\rm{M_{500}}\gtrsim 2 \times 10^{14}~M_{\odot}$\footnote{$\rm M_{500} = \frac{4}{3} \pi 500 \rho_c R_{500}^3$ with $\rho_c$ being the critical density and $R_{500}$ the radius within which the average cluster density is 500 $\rho_c$ }) and the highest-mass systems observed in the Universe (Tier-2: z$<0.6$; $\rm{M_{500}}>7.25 \times 10^{14}~M_{\odot}$).
The uniform depth of the acquired X-ray data ensures a detailed mapping of the ICM thermodynamic properties. Therefore, the CHEX-MATE cluster sample meets the ideal requirements for a systematic and statistical analysis of cluster properties, allowing studies in a wide range of masses and redshifts.
\subsubsection{X-ray data reduction}
The X-ray data reduction adopted is extensively described in \cite{Bartalucci23} and here we report only the main points.\\
Target observations have been performed with the European Photon Imaging Camera (EPIC, \citealt{turner2001} and \citealt{struder2001}) and the data were reprocessed using the Extended-Science Analysis System (E-SAS\footnote{\href{http://cosmos.esa.int/web/xmm-newton}{cosmos.esa.int/web/xmm-newton}}, \citealt{snowden08},  SAS version 16.1). To remove flare events, time intervals with count rates exceeding $ 3 \sigma$ above the average value were excised by using the tools \emph{mos-filter} and \emph{pn-filter} in the [2.5-8.5] keV energy range. 
Point sources were masked from the analysis following the scheme detailed in Section 2.2.3 of \cite{ghirardini19} and further described in \cite{Bartalucci23}.\\
Images were obtained by extracting the photon-count images in the [0.7-1.2] keV band \citep[i.e. the band that maximizes the signal-to-background of the cluster thermal emission,][]{Ettori2000} for each camera and produced the exposure maps with the tool \emph{eexpmap}.
The instrumental (particle) background has been removed by producing images accounting for both particle and residual soft protons, following the procedure described in \cite{ghirardini19}.
Briefly, the former component is estimated from the rescaled filter-wheel-closed spectra
and fitted with a model including a broken power-law and several Gaussians to reproduce fluorescence lines; the latter was accounted for by including in the background model an additional broken power-law, but only for those observations showing a significant contamination by high-energy particles (i.e. IN-OUT indicator higher than 0.1 $\rm counts ~ s^{-1}$).
Finally, the images, exposure, and background maps of the three cameras were then merged to maximise the statistics following the procedure described in \cite{Bartalucci23}.
\subsection{LoTSS}
The LOFAR Two-meter Sky Survey (LoTSS) \citep{LoTSS2017} provides deep, high resolution ($\sim 6"$) and high sensitivity ($\sim 100~\mu\rm{Jy~ beam^{-1}}$) images of the Northern sky in the  $120 - 168$ MHz range. Its first data release (LoTSS-DR1) was in 2019 \citep[][released area  $\sim 400$ $\rm{deg^2}$ ]{LoTSS2019}, the second release in 2022 \citep[][$\sim 5,700~ \rm{deg^2}$]{Shimwell2022} and recently the third data release has provided images for the 88\% of the Northern sky \citep{Shimwell2026}. 
One of the main goals of this survey is to provide a complete census of diffuse radio sources in galaxy clusters, such as giant radio halos, enabling systematic studies of their origin and providing stringent observational constraints on theoretical models.
Therefore, given its high sensitivity to diffuse sources at low frequencies, the LoTSS provides the ideal dataset to study in detail the properties of radio halos which shine brighter at the low frequencies probed by the survey.\\
For this work, we will use the public data of the LoTSS-DR2 clusters which have been described in \cite{Botteon2022}.
\subsubsection{Radio data reduction}\label{sec:radio_data}

In the following, we report an overview of the calibration and imaging procedures applied to the radio data. The complete description of the reduction process is presented in \cite{Botteon2022}.\\
LoTSS pointings are typically obtained with an integration time of 8hr and in the 120-168 MHz frequency range. The collected data were processed with fully automated pipelines developed by the LOFAR Surveys Key Science Project team in order to correct for direction-independent and direction-dependent effects \citep{vanWeeren2016,Williams2016,deGasperin2019,Tasse2021,vanWeeren2021}. 
As the main goal of this work is to study the diffuse halo emission, we exploited the lowest resolution images available, with a resolution  $\sim$100 kpc at the cluster redshift (corresponding to $\sim 20"-60"$ angular resolution in the considered redshift range) and with the further excision of compact source emission. Additionally, we performed a visual inspection checking for contaminating sources, both compact and extended, and masked them out in both X-ray and radio images. Specifically, we masked pixels displaying residual source emission as a consequence of poor subtraction, such as tailed AGN, and very low (or high) surface brightness regions not associated with the radio halo emission (e.g. diffuse emission clearly detached from the central halo, obvious radio fossil plasma sources or cluster's sub-components) \ {following what was done in the LoTSS-DR2 image presented in \citealt{Botteon2022}}\footnote{\url{https://lofar-surveys.org/planck_dr2.html}}).
\subsection{Sample selection}\label{Sec:sample}
As we aim to study the connection between radio and X-ray diffuse emission, we focus on the CHEX-MATE clusters covered by the LoTSS-DR2 that have been proved to show diffuse radio emission. Thus, of the 40 targets in common, we constructed our final sample by selecting those 18 systems classified as radio halo clusters by \cite{Botteon2022}. 
These objects have been previously analysed in \cite{Balboni2025} and we refer the reader to that work for an extensive presentation of the sample, while in the following we summarise their main properties.\\
The selected objects span a rather wide range in both redshift (0.072 - 0.575) and mass ($ 2.57 - 11.00 \times 10^{14} ~ M_{\odot}$)\footnote{All the considered masses are taken from the MMF3 Planck catalogue \cite{Planck2016}} (see Tab.~\ref{tab:sample_18}). 
They belong to both CHEX-MATE Tier-1 and Tier-2 subsamples, well sampling the whole mass and redshift range up to $z\sim0.4$, and with two targets located beyond this value.
Furthermore, these clusters exhibit a substantial diversity in their dynamical states. As shown in Fig.~1 of \cite{Balboni2025}, they cover the majority of the dynamical range of CHEX-MATE, ranging from very disturbed objects to mildly-relaxed ones. 
We note how the selected 18 clusters do not cover the extreme relaxed end of CHEX-MATE, which might be attributed to the further selection introduced by the presence of radio emission.
However, we highlight that LOFAR is capable to probe faint diffuse emission also in cluster with low dynamical activity, as shown in past literature works \citep[e.g.][]{Cassano2023,Biava2024,vanWeeren2026} and by the wide dynamical range covered by our sample. Therefore, the radio selection is significantly mitigated with respect to past radio observations \citep[e.g.][]{Cuciti21-2}.\\
In addition, as discussed in \cite{Balboni2025}, CHEX-MATE more massive systems are located at higher redshift by construction, thereby introducing a correlation between $M_{500}$ and z.
Both this and the selection due to the radio emission are two factors that we have to keep in mind when discussing our results.
\section{Data analysis and Results}\label{sec:data_analysis}

\subsection{Point-to-point analysis}\label{sec:ptp}
\cite{Govoni2001} showed the importance to study the correlation between the X-ray and radio brightness and posed the basis of its use to derive information on the mechanism responsible for radio emission. This approach has been then adopted in many subsequent studies and for different radio frequencies \citep[e.g.][]{Hoang2019,Kamlesh-M0717}.\\
Past works have shown that perturbed galaxy clusters tend to present a positive correlation between radio and X-ray emission which, in general, is found to be sub-linear \citep[e.g.][]{Giacintucci2005,Cova2019,Xie2020,Hoang2021}. Other sources have been found to have different relations. For example, mini-halos typically displayed a super-linear $I_R-I_X$ trend (e.g. \citealt{Ignesti-2020-MH}) while relics have been proved to present an anticorrelation between the two bands \citep[e.g.][]{Campitiello2024}.\\
However, as also presented in several studies \citep{Botteon2020-A2255,Bonafede2022,Kamlesh2021-A2744,Sikhosana23}, differences in the instrument, observing frequency, angular resolution, and the adopted analysis technique, can impact the final results, making difficult to draw general conclusions.\\
Thanks to LoTSS and CHEX-MATE high-sensitivity images, we can now overcome these limitations and consistently compare the results obtained for the radio and X-ray relations in different clusters.\\
To perform such a study on our sample, we extract the average surface brightness from radio ($I_R$) and X-ray ($I_X$) images by constructing a grid \citep[following e.g.][]{Govoni2001,Kamlesh-M0717} based on the radio halo emission, which, in general, appears less extended than the X-ray one. Each grid cell has an angular size equal to the radio beam resolution, which is typically larger than the XMM-Newton point spread function in the X-rays ($\sim 15''$). 
Following \cite{Botteon2020-A2255}, we consider only the cells where the average radio surface brightness is above $2 \times \sigma_{\rm RMS}$ (see also Appendix~\ref{appendix:test_ptp}). 
Prior to extracting the values from the grid, as detailed in Sec.~\ref{sec:radio_data}, we took particular care in the excision of sources which are unrelated to the diffuse radio halo emission.\\
Once the grid is set, we proceeded with the surface brightness extraction in each box from both radio and X-ray images. In particular, we computed the cluster count rate in the 0.7-1.2 keV energy band from the background-subtracted and exposure-corrected XMM images and then converted them into physical units as well as accounting for the $k$-correction and cosmological dimming as explained in \cite{Balboni2024}.\\
We have investigated the thermal -- non-thermal relation fitting $I_X$ and $I_R$ with a power law in log-space :
\begin{equation}
    {\rm log} I_R = {\rm A} ~  {\rm log} I_X + {\rm N},
\end{equation}
where the slope A tells us whether the radio component (CRe and magnetic fields) distribution is steeper (A$>1$) or flatter (A$<1$) than the thermal ICM distribution.\\
To perform the fit, we used a Bayesian regression model which accounts for x and y errors, selection effects (i.e. Malmquist bias, \citealt{Malmquist1922}) and the intrinsic scatter ($\sigma_{int}$) in the regression relationship \citep[see][for further details]{Balboni2024}.
\begin{figure*}[ht]
    \centering
    \includegraphics[scale=0.2]{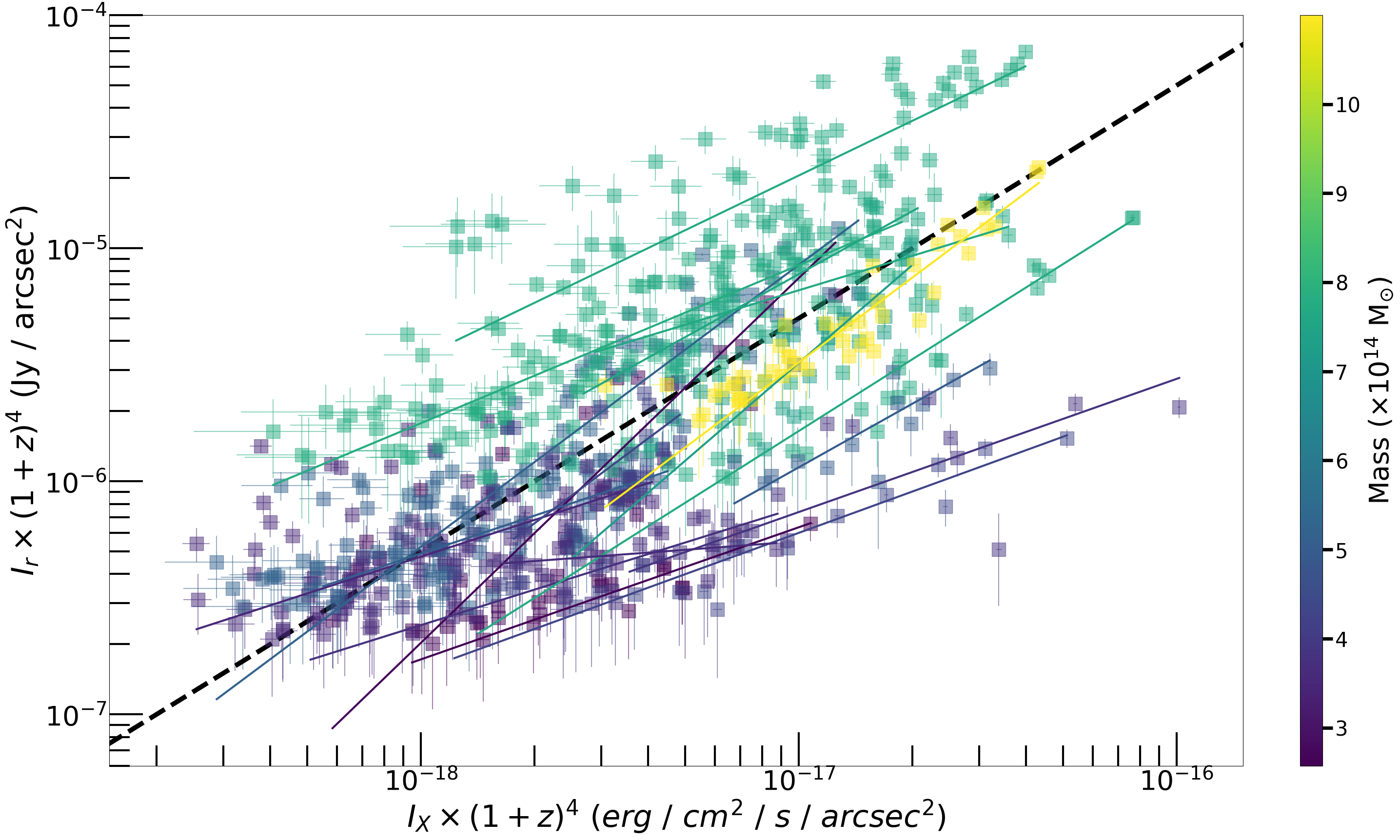}
    \caption{$I_R - I_X$ relations of all the clusters, color-coded by the cluster mass. Solid lines show the best-fit relations while the dashed black line highlights a linear relation for reference.}
    \label{fig:Ix-Ir_tot}
\end{figure*}
Fig.~\ref{fig:Ix-Ir_tot} shows the point-to-point comparisons between the radio brightness at 144 MHz and the X-ray brightness in the range 0.7-1.2 keV for our sample.\\
\begin{table}[]
    \centering
    \caption{Spearman correlation coefficients, correlation slopes and intrinsic scatter of the $I_R-I_X$ relations.}
    \label{tab:2}
    \begin{tabular}{lrrrrr}
    \toprule
    Name & $r_S$ & A$\pm \delta A$ & $\sigma_{int}$ \\
    \midrule
    PSZ2G031.93+78.71 & 0.75 & $1.56 \pm 0.21$ & $0.66 \pm 0.11$ \\
    PSZ2G040.58+77.12 & 0.84 & $0.57 \pm 0.14$ & $0.09 \pm 0.07$ \\
    PSZ2G046.88+56.48 & 0.56 & $0.54 \pm 0.14$ & $0.08 \pm 0.06$ \\
    PSZ2G048.10+57.16 & 0.53 & $0.52 \pm 0.10$ & $0.47 \pm 0.05$ \\
    PSZ2G049.32+44.37 & 0.53 & $0.50 \pm 0.14$ & $0.08 \pm 0.06$ \\
    PSZ2G053.53+59.52 & 0.85 & $1.21 \pm 0.08$ & $0.51 \pm 0.05$ \\
    PSZ2G055.59+31.85 & 0.75 & $1.02 \pm 0.06$ & $0.09 \pm 0.06$ \\
    PSZ2G056.77+36.32 & 0.73 & $0.59 \pm 0.07$ & $0.10 \pm 0.07$ \\
    PSZ2G066.41+27.03 & 0.74 & $0.77 \pm 0.11$ & $0.41 \pm 0.05$ \\
    PSZ2G077.90-26.63 & 0.80 & $0.90 \pm 0.23$ & $0.12 \pm 0.10$ \\
    PSZ2G083.29-31.03 & 0.48 & $0.48 \pm 0.18$ & $0.52 \pm 0.10$ \\
    PSZ2G107.10+65.32 & 0.86 & $0.69 \pm 0.05$ & $0.29 \pm 0.04$ \\
    PSZ2G111.75+70.37 & 0.47 & $1.20 \pm 0.46$ & $0.39 \pm 0.09$ \\
    PSZ2G113.91-37.01 & 0.77 & $0.90 \pm 0.15$ & $0.32 \pm 0.06$ \\
    PSZ2G143.26+65.24 & 0.90 & $1.40 \pm 0.21$ & $0.23 \pm 0.06$ \\
    PSZ2G179.09+60.12 & 0.46 & $0.57 \pm 0.22$ & $0.56 \pm 0.16$ \\
    PSZ2G186.37+37.26 & 0.94 & $1.21 \pm 0.09$ & $0.18 \pm 0.04$ \\
    PSZ2G192.18+56.12 & 0.05 & $0.11 \pm 0.23$ & $0.15 \pm 0.11$ \\
    \bottomrule
    \end{tabular}
\end{table}
We observe a clear correlation between X-ray and radio brightness in every target. The only exception is PSZ2G192.18+56.12 displaying a Spearman correlation coefficient of $r_S = 0.05$.
Fig.~\ref{fig:Ix-Ir_tot} also shows that, when significant, a positive $I_R - I_X$ correlation exists. We report in Tab.~\ref{tab:2} the best-fit values for A and $\sigma_{int}$ of the linear relation\footnote{We note that the A values of the five objects presented in \cite{Balboni2024} are different from the values reported here. This is due to: (i) a different treatment of the 2$\sigma_{RMS}$ threshold in the fitting process, which now accounts also for the k-correction applied to the cluster $I_R$; (ii) a more conservative calculation of $\sigma_{RMS}$ computed automatically for all the clusters; (iii) the exclusion of the ``ridge'' region in PSZ2G046.88+56.48. The final values are consistent within the errors and do not affect neither this nor previous results.}.\\
As in \cite{Balboni2024}, we investigated whether breaks in the single power-law were present by fitting the $I_R-I_X$ relation with a broken power-law. Excluding the cases already discussed in \cite{Balboni2024}, no clear statistical evidence was found for preferring a broken power-law function in any targets. Hence, a single power-law is already providing a good representation of the observed point-to-point relation of these clusters.
\subsection{Radio - X-ray displacements}\label{sec:R-X_offset}
\begin{figure}
    \centering
    \includegraphics[width=0.49\linewidth, height=5cm]{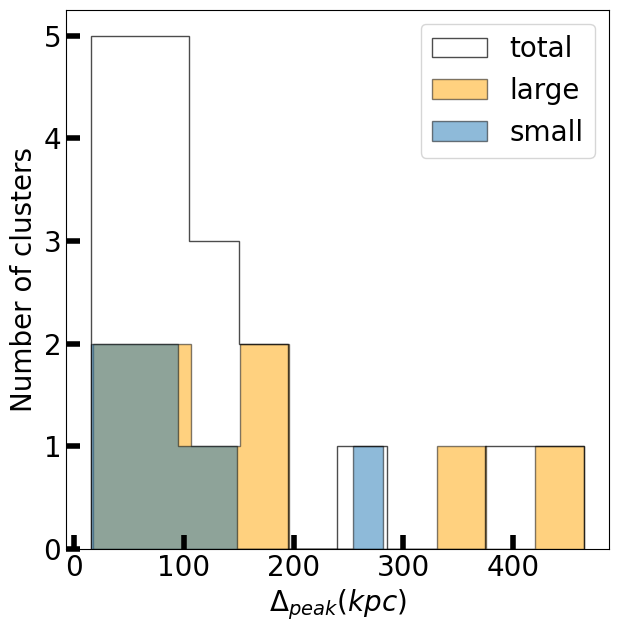}
    \includegraphics[width=0.49\linewidth, height=5cm]{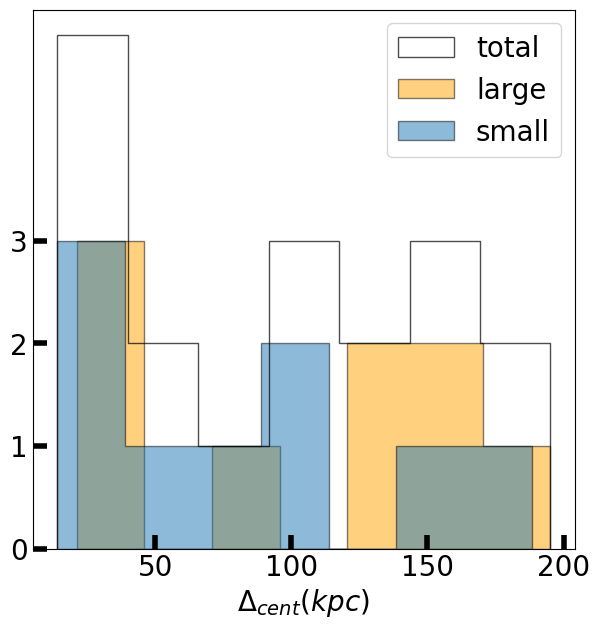}
    \caption{Histograms showing the peak (left) and centroid (right) separation between radio and X-ray emission. Different colors identify subsamples considering halos with size larger (orange) or smaller (blue) than the median of our sample (660 kpc). }
    \label{fig:hist_distance}
\end{figure}

\begin{figure}
    \centering
    \includegraphics[width=0.8\linewidth]{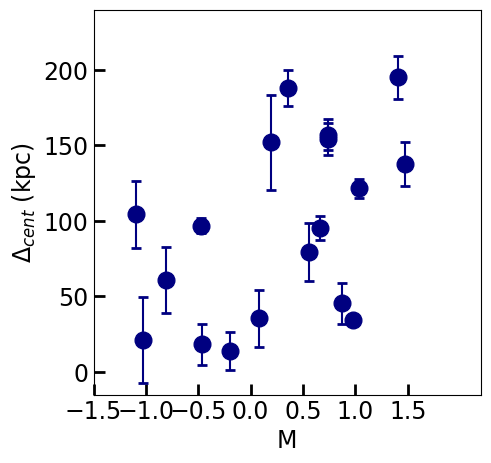}
    \caption{Relation between the radio and X-ray centroid offsets and the cluster dynamical disturbance (M) for the 18 objects analysed here.}
    \label{fig:cent_dist-M}
\end{figure}

As one of the aims of this work is to study the radial variation of the radio and X-ray relation (see Sec.~\ref{sec:radial_slopes}), we need to define the central point of the radial profile, which will be common for both radio and X-ray maps.
If the radio and X-ray diffuse emission were perfectly coincident, the choice of such a point would be straightforward. However, as already pointed out by \cite{Feretti2012}, this may not always be the case, in particular when resolving in details the radio emission.
Therefore, it is worth to investigate the presence of displacement between the radio and X-ray emissions. 
In addition, despite the impact of the profile centre choice has been explored in past X-ray analyses \citep[see. e.g.][]{Bartalucci23}, no study of such kind has been made for the radio component.
\cite{Feretti2012} compared the separation between the radio halo centre and the centroid of the X-ray emission. They reported offsets up to hundreds of kpc in both large ($\geq 1$Mpc) and small ($<1$Mpc) radio halos, with the latter showing more pronounced displacements with respect to their size.
To investigate such features in our sample, we consider the peak and the centroid of both the thermal ICM and radio halo emission. 
The results showing the separation between radio and X-ray emission peak and centroid are reported in Fig.~\ref{fig:hist_distance}.
We also defined small and large halos based on their median size ($\sim 660$ kpc, as defined in \citealt{Balboni2025}).
From Fig.~\ref{fig:hist_distance} we see that, typically, there is a non-negligible offset between the radio and X-ray emission, either considering the peak or the centroid of the two.
Despite the sparse measurements, centroid displacements are, on average, smaller than the peak ones. This can be expected as the centroid is taking into account the whole morphology of the emission which, as we demonstrated also from the point-to-point analysis, is, on average, spatially correlated between the two bands.\\
In addition, given the wealth of X-ray information available for CHEX-MATE clusters, we were able to find a correlation between the radio and X-ray offset with the cluster dynamical state (see Fig.~\ref{fig:cent_dist-M}). The latter is quantified by the M parameter, which combines a number of morphological indicators presented in \cite{Campitiello2022} by summing the deviations of each parameter from the mean of its distribution in units of standard deviation \citep{Rasia2013}. Large and small values of M identify more and less dynamically disturbed clusters, respectively. 
The observed correlation ($r_{\rm S}\sim0.48$, $p=0.04$) suggests that the radio--X-ray centroid offset may serve as an additional tracer of the cluster dynamical state. A similar, although weaker, correlation is  also found for the peak offset ($r_{\rm S}\sim0.4$).\\
Therefore, given that (i) centroid estimates take into account the whole emission morphology in both bands and (ii) centroid offsets between radio and X-ray component present, on average, lower values, we will consider as profile centre the midpoint between the radio and X-ray centroids.
In this way we will further mitigate the effects of those clusters with large radio-X-ray offsets.
\subsection{Radial analysis}\label{sec:radial_slopes}
Past works have shown how the radial analysis of thermal and non-thermal properties provides additional information about the ICM environment \citep[e.g. interaction between different ICM components and on the properties of CRe and magnetic fields at large radii;][]{Bruno2021, Cuciti22}. 
As done in the pilot study of \cite{Balboni2024}, we investigated the radial trends of the $I_R - I_X$ slope as well. 
The value of the slope tells us how radio and X-ray components relate, and finding a varying (or constant) relation would imply a change (or not) also in the relative contribution of the two plasma components. 
We derive the radial slope as:
\begin{equation} \label{eq:k}
    k(r) = \frac{\Delta (\log I_R)}{\Delta (\log I_X)},
\end{equation}
where $k(r)$ is the correlation slope of the $I_R \propto I_X^k$ scaling and $\Delta \log I_R$ ($I_X$) is the difference between the logarithm of the radio (X-ray) brightness in two consecutive annuli. 
To balance between a sufficiently high signal in each bin and a good radial resolution, we computed the radial slope using two different widths of the annuli, specifically 1 and 2 times the beam width (which also reduces the dependence on the chosen binning, which effect is further discussed in Appendix~\ref{appendix:test_binning}  together with possible systematics due to the limited sample size). We note that only for three objects we were not able to compute the profiles using the minimum distance binning. 
\begin{figure}[h!]
    \centering
    \includegraphics[width=\linewidth]{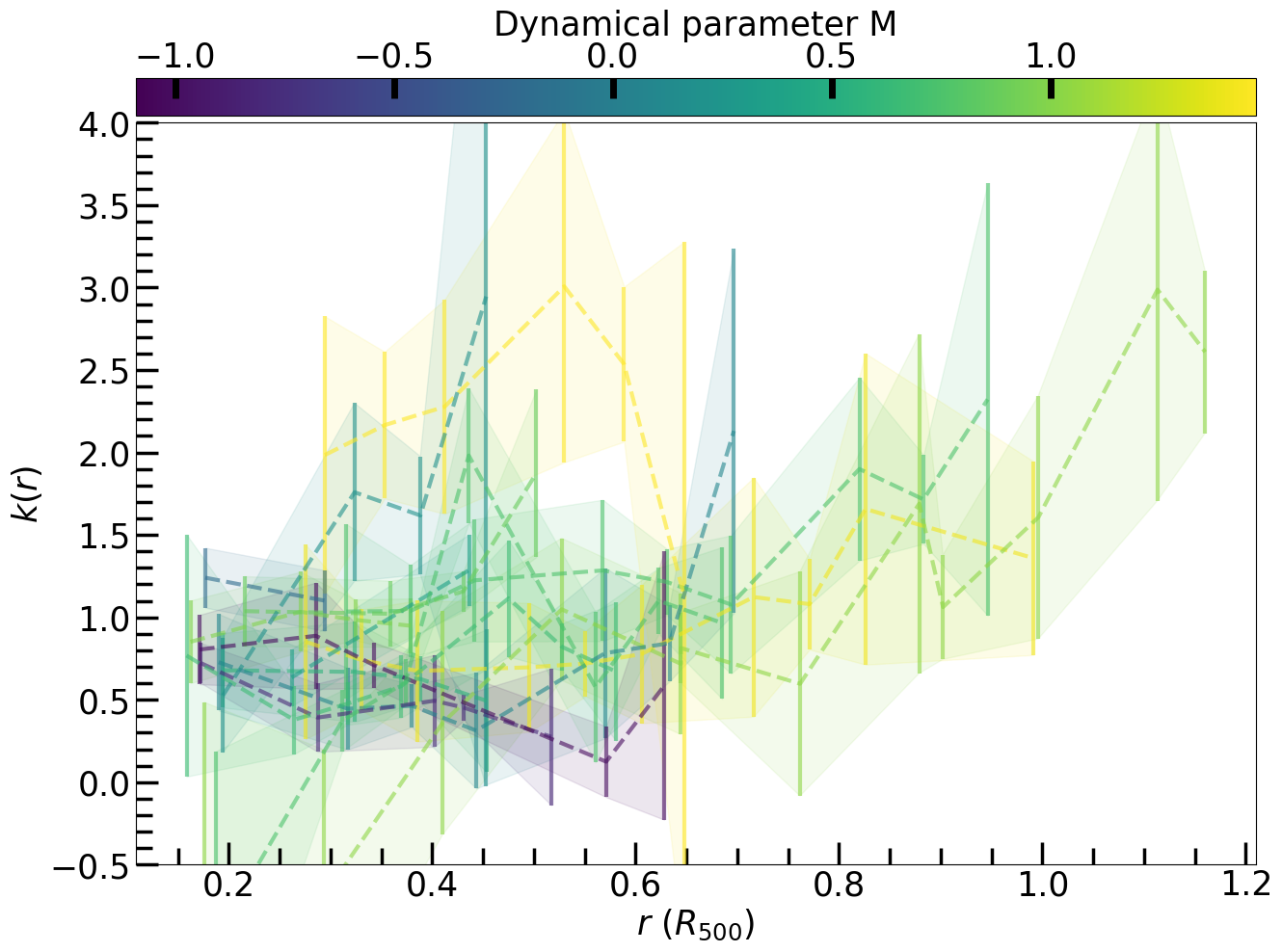}
    \caption{Radial profile of the radio-X slope ($k(r)$) for the studied clusters colour coded by their M parameter.}
    \label{fig:slope_trends}
\end{figure}
Fig.~\ref{fig:slope_trends} shows the evident complex trends in the presented halos, displaying how a uniform $k$ is generally not representative of the whole cluster extension.
\section{Discussion}\label{sec:discussion}
\subsection{Comments on point-to-point analysis}\label{sec:disc_ptp}
\begin{figure}[ht]
    \centering
    \includegraphics[width=\linewidth]{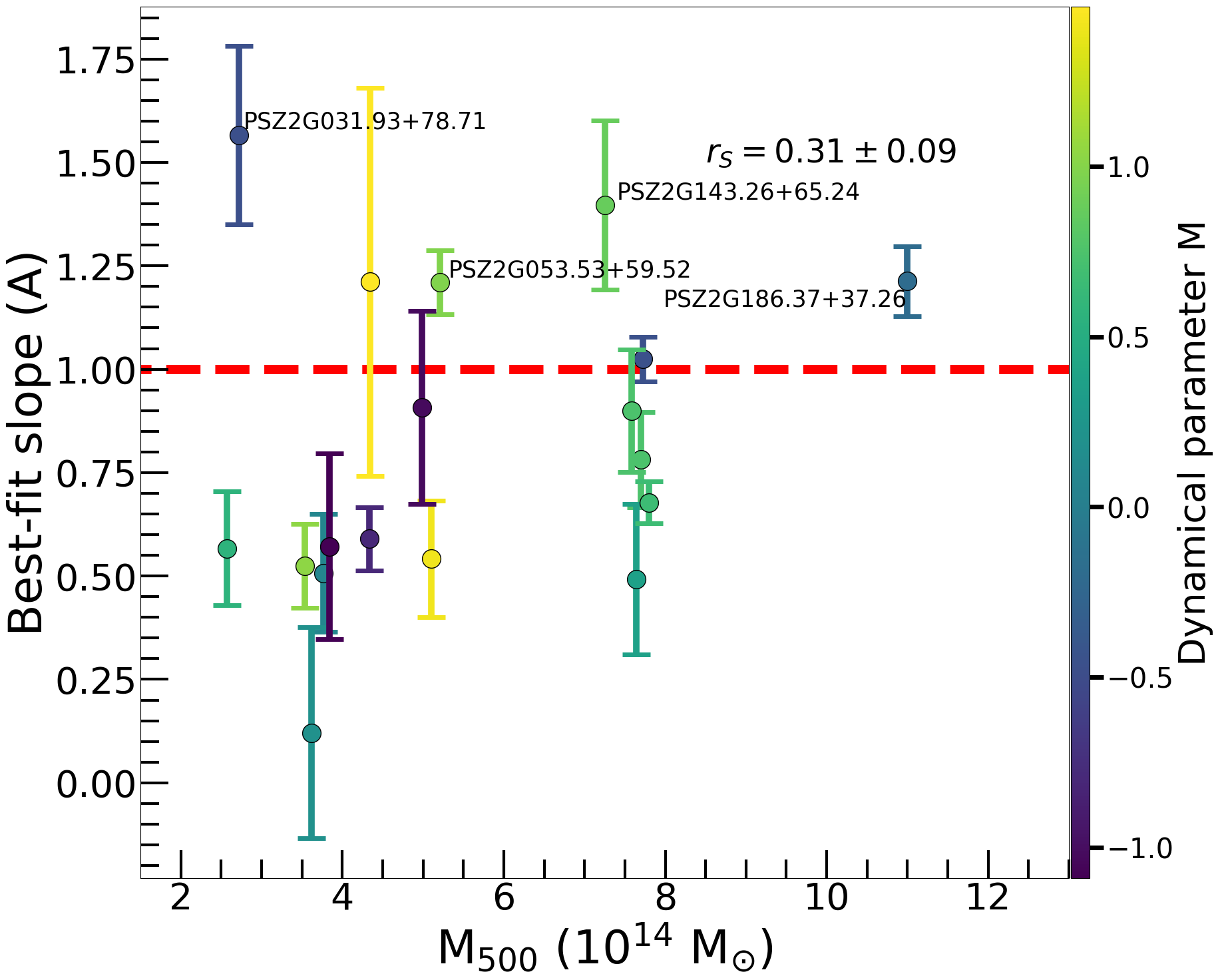}
    \caption{Comparison between $M_{500}$ and the best-fit slope in our sample, with highlighted the cluster ID for those targets showing superlinear slopes.}
    \label{fig:mass_cmd-slope}
\end{figure}

\begin{figure}
    \centering
    \includegraphics[scale=0.13]{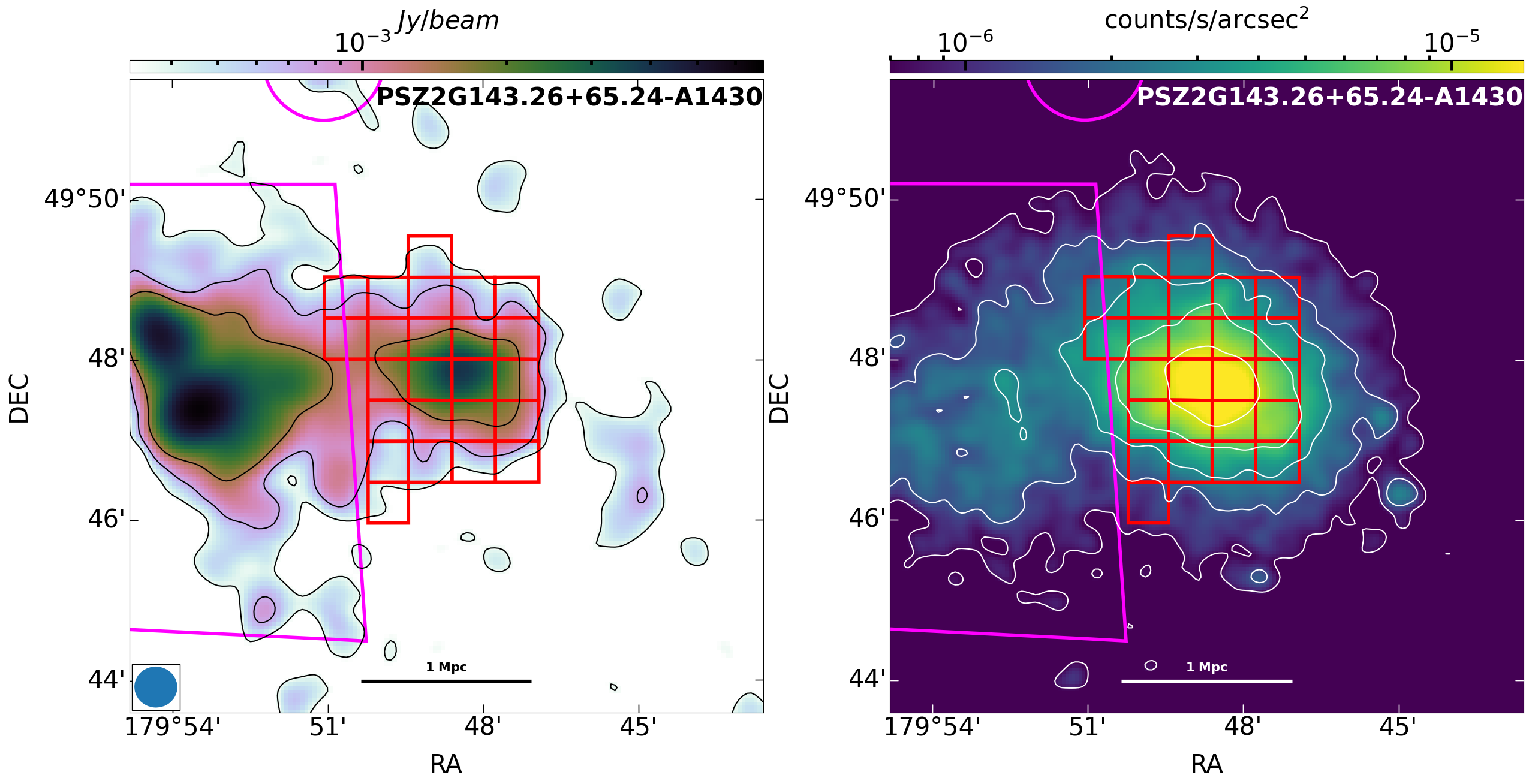}
    \includegraphics[scale=0.13]{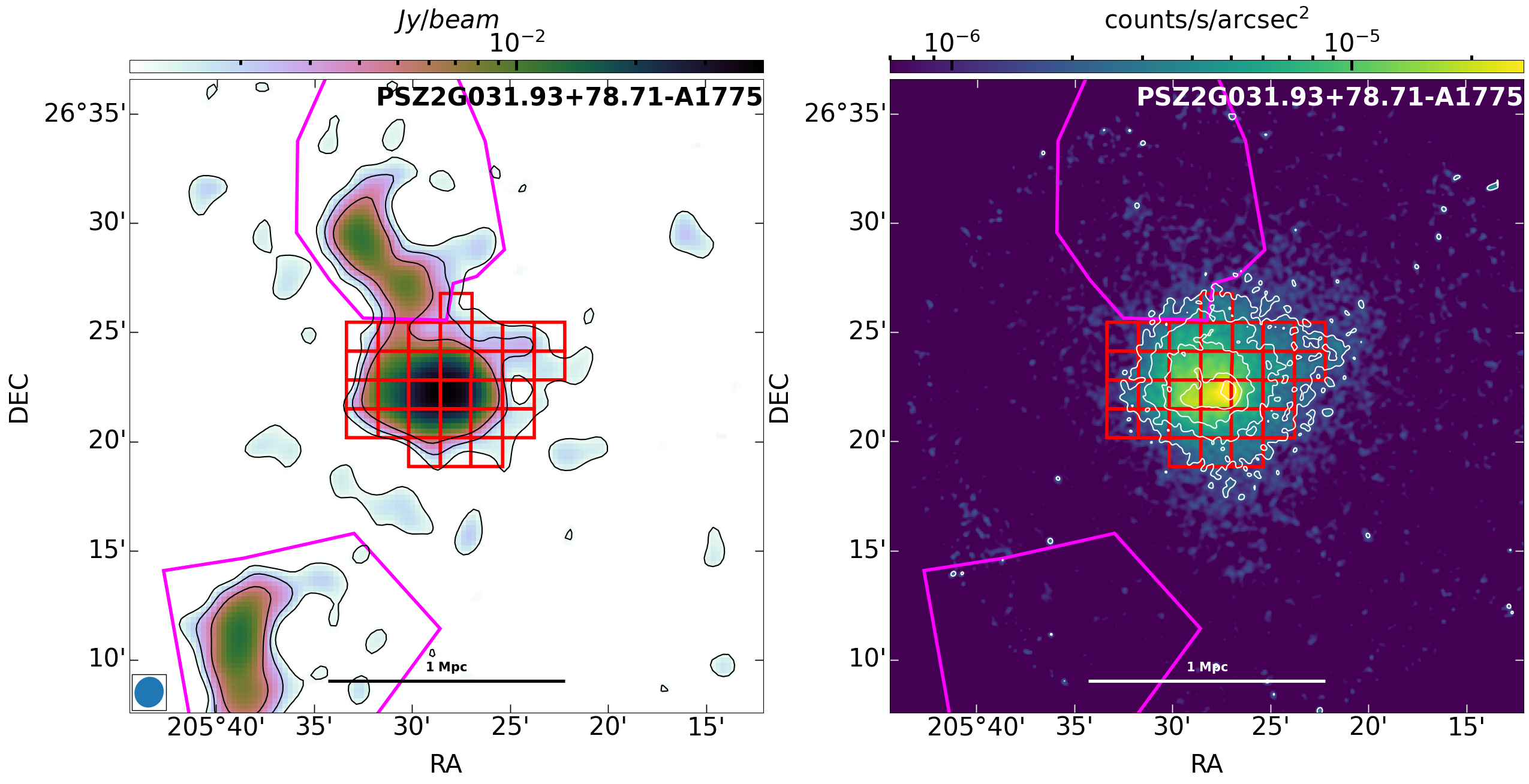}
    \includegraphics[scale=0.13]{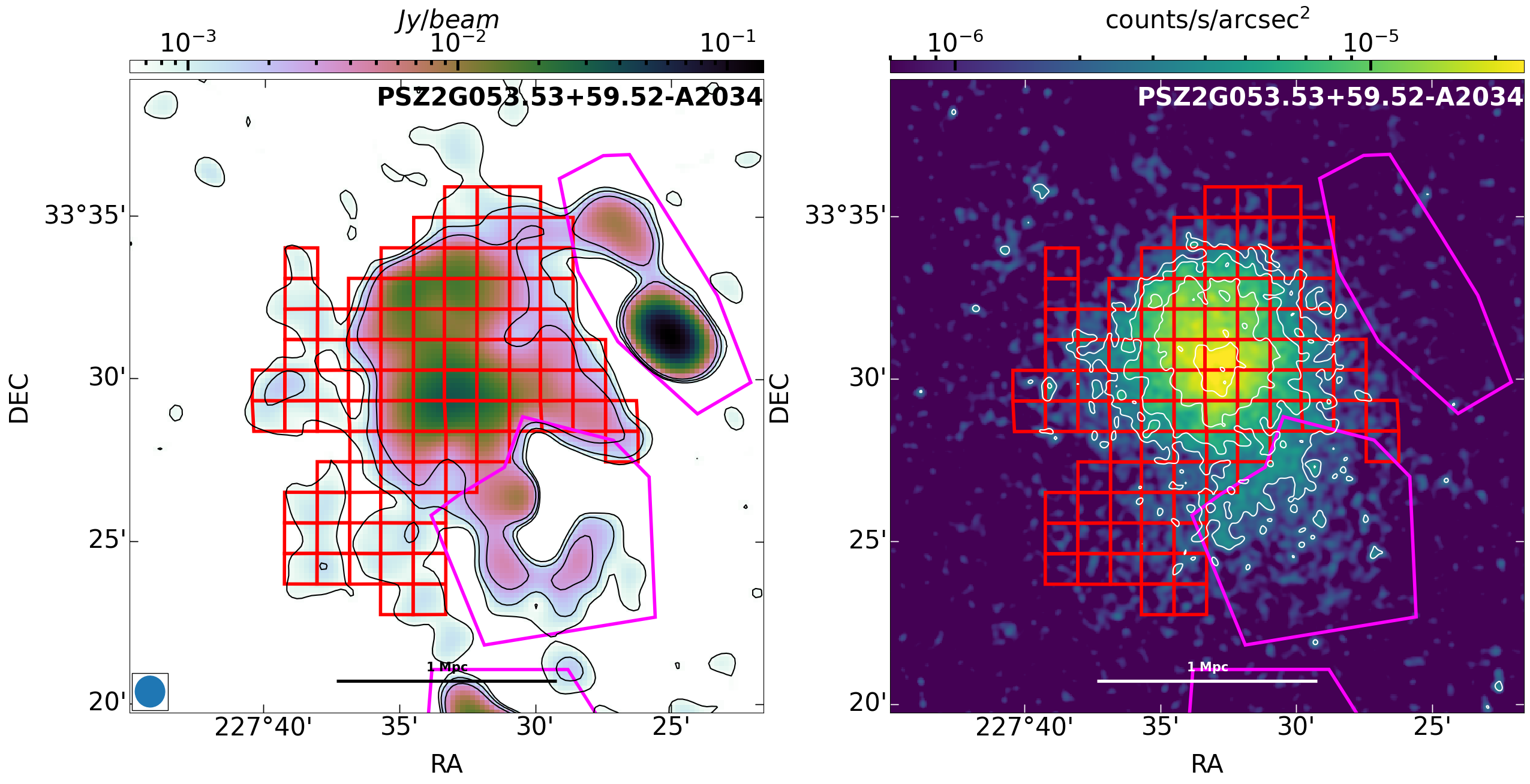}
    \includegraphics[scale=0.13]{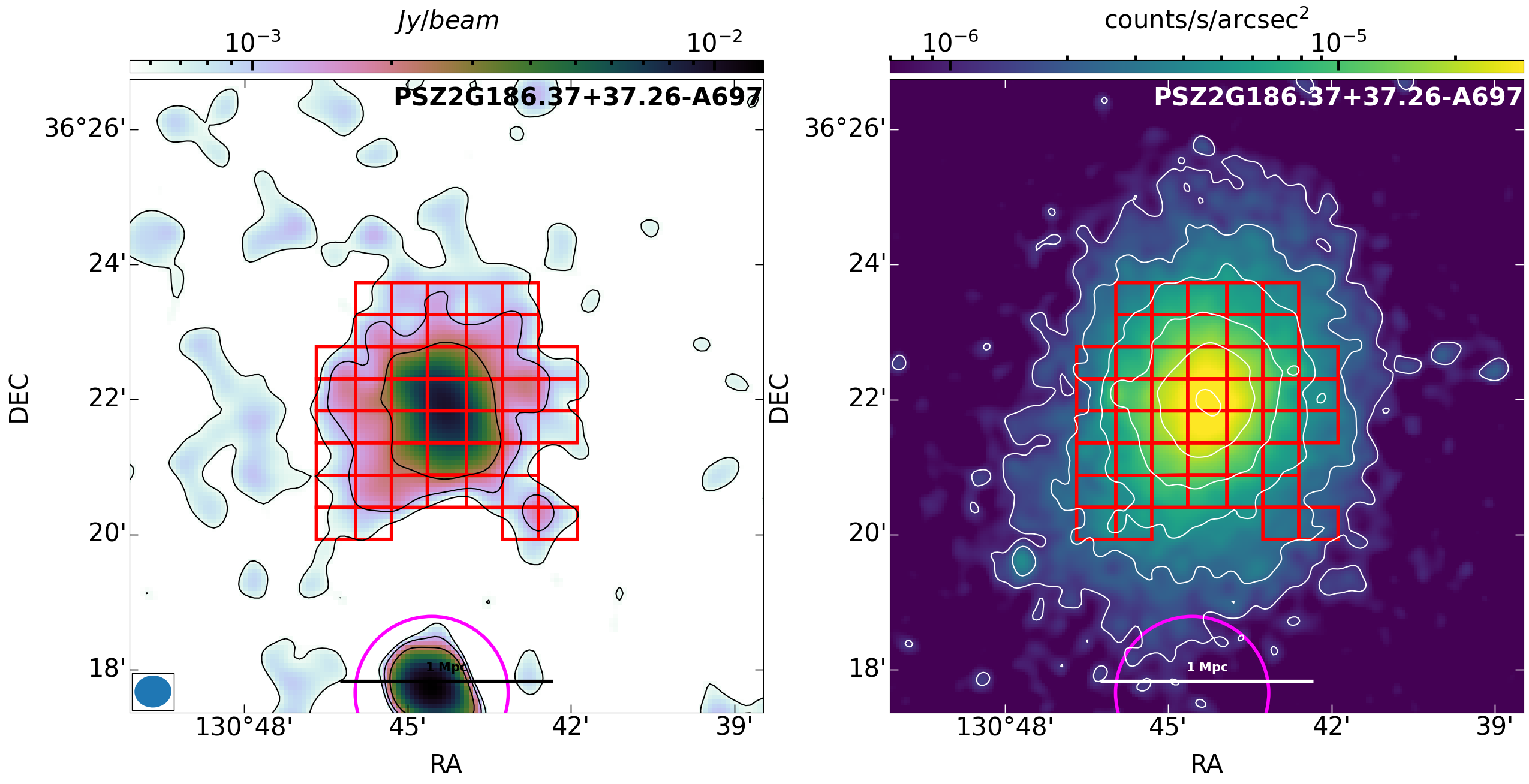}
    \caption{Radio and X-ray images of the four targets with superlinear $I_R - I_X$ slopes. The red boxes indicates where we extracted the surface brightness, while the magenta regions are the excluded ones. The radio contours are plotted at $2,4,8 \times {\rm \sigma_{RMS}}$, while the X-ray ones start from the average value of the background subtracted and exposure corrected map and are spaced by a factor of $2$. }
    \label{fig:suplin_clusters}
\end{figure}

\begin{figure}
    \centering
    \includegraphics[width=\linewidth]{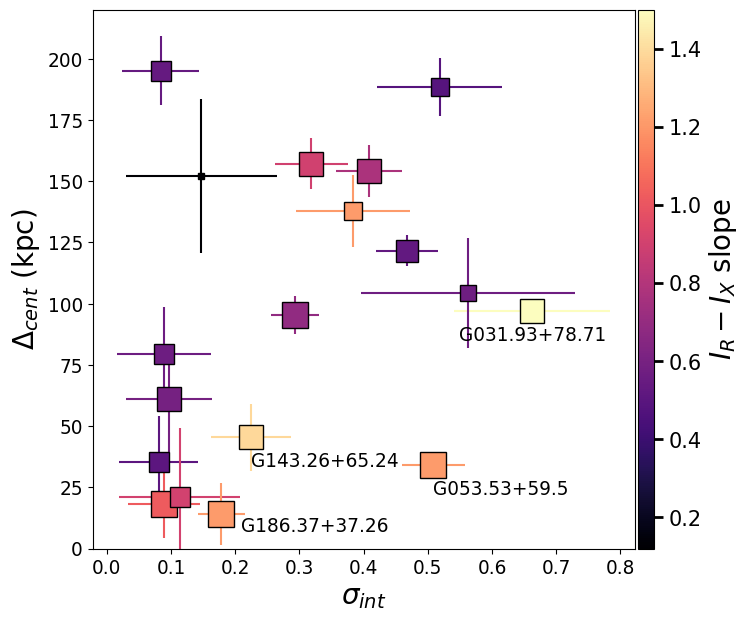}
    \caption{Comparison between the intrinsic scatter of the point-to-point relation and the centroid separation ($\Delta_{cent}$). The points are colour-coded according to A and their size is proportional to slope measurement accuracy ($\rm \Delta A/A$).}
    \label{fig:scat_deltacent_slope}
\end{figure}

Figure~\ref{fig:Ix-Ir_tot} shows the tight correlation among the thermal and non-thermal components in the clusters of our sample.
\ {In particular, in the upper panel one can see the clear mass and/or redshift dependence, where brighter radio halo clusters are also the more massive ones and located at higher redshifts. This is in line with what \cite{Cuciti2023} found for the whole Planck-LoTSS-DR2 sample and previous observational findings.
As we are focusing on the cluster surface brightness emission}, in Fig.~\ref{fig:Ix-Ir_tot_cpar} we also report the same relations but colour-coded each cluster according to the dynamical state-related parameter $c$ (light concentration, \citealt{Campitiello2022}). This parameter quantifies the amount of X-ray emission within the cluster centre ($0.15 R_{500}$ in this case) with respect to the total one within $R_{500}$. Therefore, more dynamically relaxed clusters, exhibiting a central X-ray peak, would display higher values of $c$ compared to more disturbed ones. 
In our sample, we find that objects with different dynamical states are located in different regions of the plot.
In particular, more concentrated clusters (see also e.g. \citealt{Cassano2010a,Cuciti21-2}) have fainter radio halos but brighter X-ray emission, while bright radio halos display lower values of $c$.\\
In Figure~\ref{fig:Ix-Ir_tot}, we also report the best-fit power-law relation found for each cluster.
As pointed out by \cite{Govoni2001}, the point-to-point analysis gives not only information on the spatial distribution of the thermal and non-thermal ICM, \ {but it also provides useful observational tests for the mechanism responsible for halo emission.}
They showed how, under some reasonable assumptions on the magnetic field, the expected correlation between $I_X$ and $I_R$ in (re-)acceleration and hadronic models is different. In particular, the sublinear scaling often observed in radio halos is hardly reproduced by the hadronic scenarios, as they would require an unphysical amount of CR protons in the cluster peripheries to justify such $I_R-I_X$ relation \citep[e.g.][]{Brunetti14}.\\
To investigate this behaviour, from the posterior distributions of each fit, we computed the probability $P({\rm A}<1)$ for each target. We then estimated the ensemble fraction of sub-linear systems via Monte Carlo resampling of the individual posteriors, obtaining $67 \pm 5 \%$ of the sample showing sub-linear $I_R - I_X$ relations.
Although sublinear radio--X-ray relations have been reported in a few individual clusters, this is the first systematic evidence that such a behaviour is widespread among radio halos. This suggests that, on average, the non-thermal component is distributed more broadly than the thermal ICM.
To help visualise this, in Fig.~\ref{fig:mass_cmd-slope} we plot A vs $M_{500}$ colour coded by M.
We can further speculate on this result considering, as a reference, the case where the magnetic field energy scales linearly with the thermal energy of the ICM, i.e. $B \propto n_{th}^{0.5}$, where $n_{th}$ is the ICM thermal gas density, as observed by \cite{Bonafede2010}.
In these conditions, a sub-linear slope of the $I_R-I_X$ relation indicates a much flatter distribution of the CRe density than $n_{th}$ (see the discussion in \citealt{Govoni2001}). This may also imply that CRes could extend much more than the detectable radio emission, but, due to the limited sensitivity, they are generally not observed (\citealt{Vazza2024}).\\
We also report four cases of clear superlinear $I_R-I_X$ trends, namely: PSZ2G031.93+78.71, PSZ2G053.53+59.52, PSZ2G143.26+65.24 and PSZ2G186.37+37.26. In these targets, we find the cluster morphology in the radio and X-ray emissions to be remarkably similar, in particular for what regards the central core of the emissions (Fig.~\ref{fig:suplin_clusters}).
Additionally, three of these targets (PSZ2G053.53+59.52, PSZ2G143.26+65.24 and PSZ2G186.37+37.26) have a small projected separation between the centroid of the two emissions with respect to the whole sample (see Fig.~\ref{fig:scat_deltacent_slope}).
PSZ2G031.93+78.71, instead, displays the highest $\sigma_{int}$ of the sample (Fig.~\ref{fig:scat_deltacent_slope}). This last feature, together with the observed $I_R-I_X$ result, resembles the case of Abell 1351 reported by \cite{Srikanth2026} for which a complex spectral index behaviour was observed. \ {In addition, both objects present bright substructures and filaments in the central part of the halo \citep{vanWeeren2026, Srikanth2026}, which can affect the resulting $I_R - I_X$ relation.}
One possible interpretation is that these systems may host a more centrally concentrated non-thermal component, linked to AGN fossil plasma or remnant mini-halo-like emission, in addition to a more extended halo one, although with the present data a break in the $I_R-I_X$ relation is not observed. 
Indeed for PSZ2G031.93+78.71, a two-components radio emission was observed by \cite{vanWeeren2026} by studying its radial profile.
These objects may therefore provide a useful bridge between giant radio halos and core-regulated non-thermal plasma.
Possible, additional tests on the presence of (remnant)cool-core signatures, central entropy drops, AGN radio activity or X-ray cavities, will help to shed light on this scenario\citep[e.g.][]{Yang2019, Gaspari2020, Cammelli2026}.\\
That said, in these targets the projected radio emission is more peaked than the thermal one. In the aforementioned framework, this evidence could be interpreted as result of a CRe radial distribution that closely follows the thermal one.
Alternatively, a tighter relation between the magnetic field and $n_{th}$ ($>0.5$) could also be the cause of the observed trends.\\
In Fig.~\ref{fig:mass_cmd-slope} we also study possible connections of the radio and X-ray relation with the mass or the dynamical disturbance, (i.e. the cluster M parameter). 
Figure~\ref{fig:mass_cmd-slope} shows how the dynamical state is unrelated to A (analogous results are found when considering $c$). 
On the contrary, the slopes could be related to the cluster mass. 
The relation has a very large scatter and the $r_S$ is low ($\sim 0.31$) and with a $p-value\sim0.08$. 
Moreover, due to the degeneracy between the cluster mass and redshift we cannot excluded the presence of evolutionary effects. Therefore, it is difficult to definitely state whether such a relation holds.
We further note that, when excluding objects with a high scatter in the $I_R-I_X$ relation ($\sigma_{int} >0.5$), the $r_S$ between A and $M_{500}$ increases to $\sim 0.65$ (p-value = 0.09).
Nevertheless, this is the first time where such a tentative correlation is reported, and future studies on larger cluster samples will be able to shed light on this connection.
\subsection{Radial slope's trend}\label{sec:disc_radial_slopes}

\begin{figure}[ht!]
    \centering
    \includegraphics[width=\linewidth]{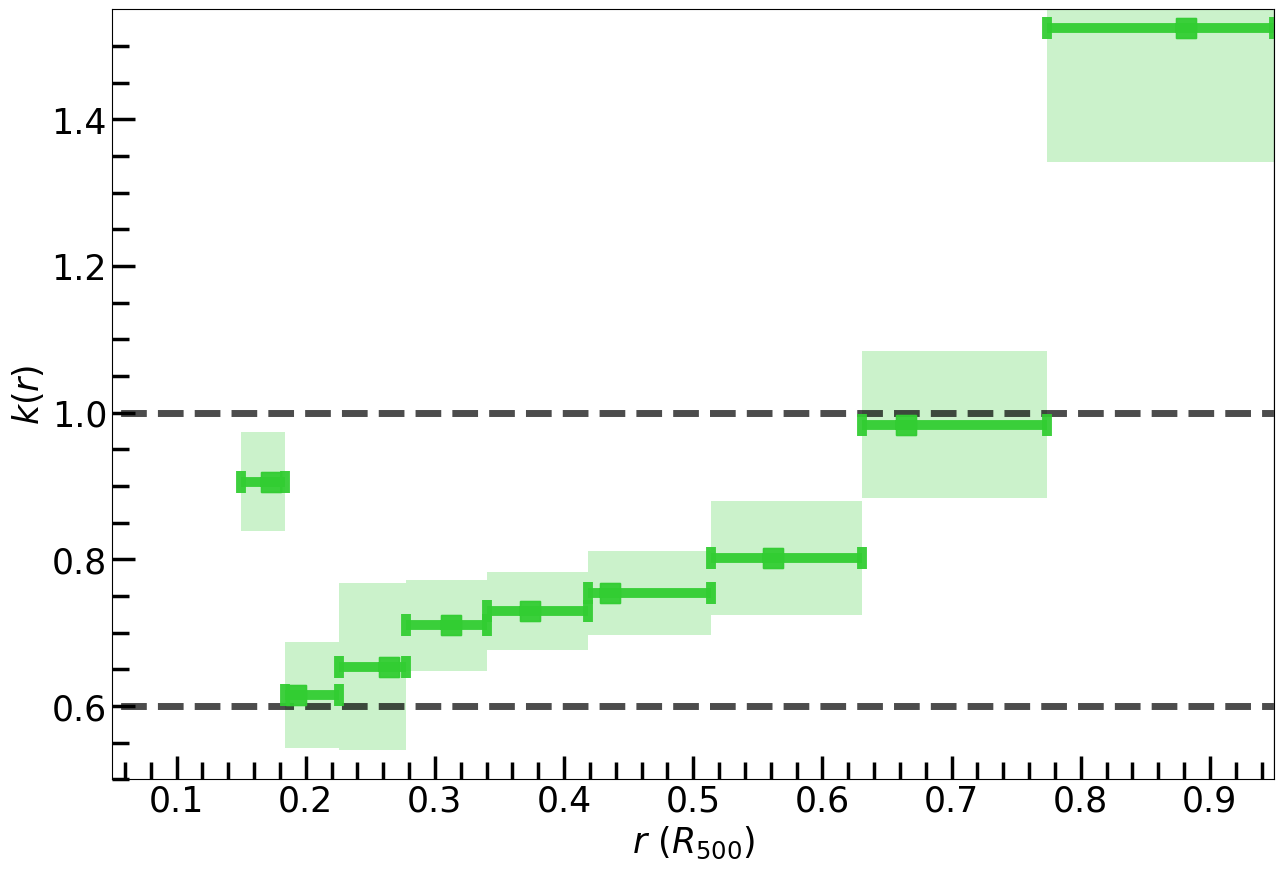}
    \includegraphics[width=\linewidth]{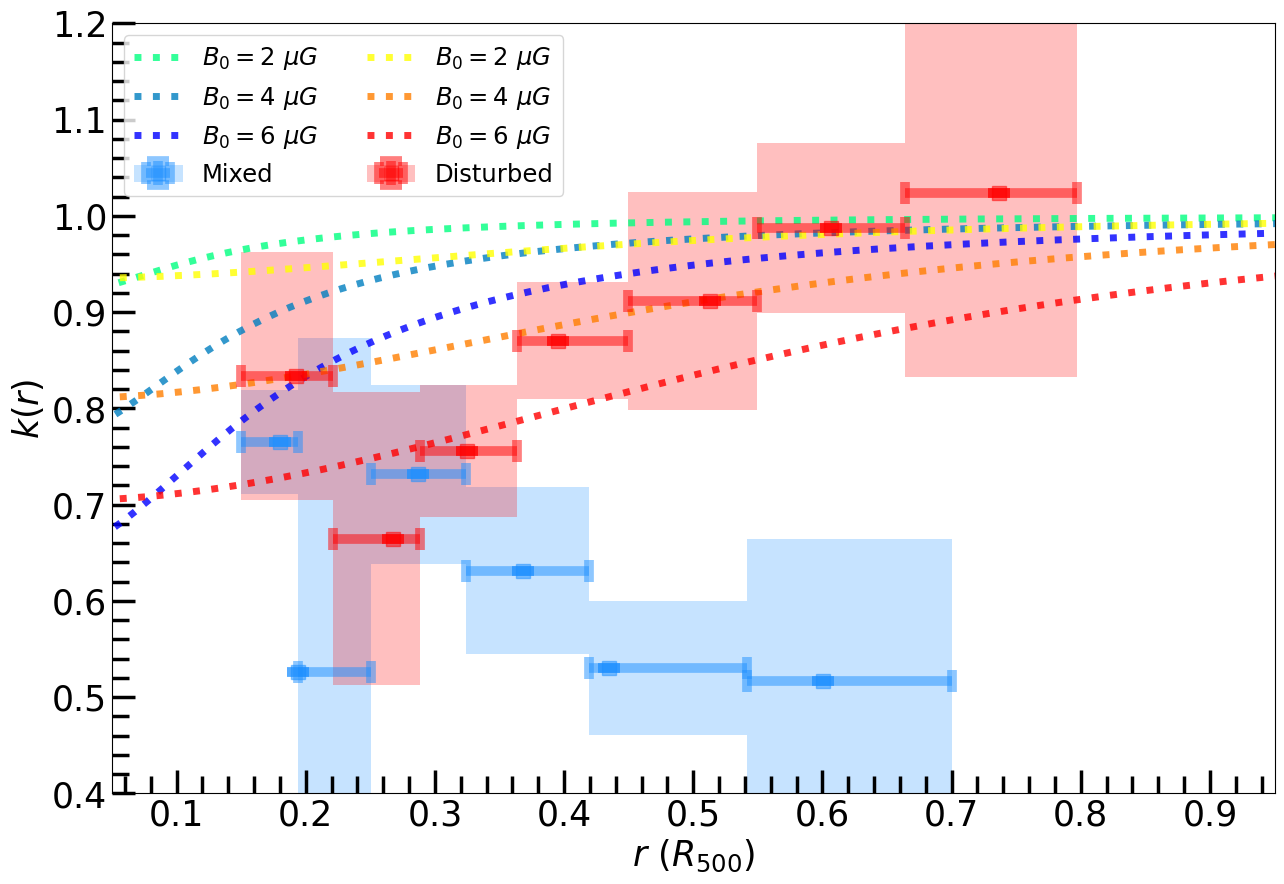}
    \caption{Weighted average profile of the $I_R - I_X$ relation slope ($k$) and its associated error (shaded area). Top: Total sample averaging, highlighting the $k = 0.6,1$ cases. Bottom: Average $k(r)$ profiles separating the cases of mixed and disturbed clusters, and the corresponding expected model prediction (Eq.~\ref{eq:em_radio}, dotted lines) for different values of $B_0$. For visualisation purposes the point at $k\sim1.5$ shown in the top panel is excluded from the plot and would belong to the disturbed sub-sample.}
    \label{fig:stacked_prof+model}
\end{figure}

Figure~\ref{fig:slope_trends} shows the wide range of $I_R-I_X$ relations found in our sample.
Despite the variety of cases, from the colour coding of Figure~\ref{fig:slope_trends} one might already identify a different behaviour between more and less dynamically perturbed objects. Specifically, the latter show hints of an increasing radial trend, especially in the very external regions.\\
In order to draw more robust conclusions, we divided the radial range in bins and computed the average of all the $k(r)$ values within each bin weighted by the inverse of the uncertainties squared.
The result is reported in the upper panel of Fig.~\ref{fig:stacked_prof+model}. However, both from Fig.~\ref{fig:slope_trends} and \cite{Balboni2024} we have evidence of differences in the $k(r)$ trends based on the dynamics of the cluster. Thus, we computed the $k(r)$ average profile also for two clusters subsamples (bottom panel) of mixed and disturbed targets.
The division is done following \cite{Campitiello2022} who classified the whole CHEX-MATE sample in Relaxed, Mixed and Disturbed systems based on their X-ray appearance. We applied the same criteria as \cite{Campitiello2022} to discriminate between mixed and disturbed clusters\footnote{We note that there are no clusters classified as relaxed in our sample.}, namely objects with M below or above 0.4. The result is significantly different from the global one, suggesting that a clear distinction of the $k(r)$ profile exists among systems with different dynamical states, and that the global profile mainly reflects the combination of these two behaviours. Therefore, we will refer to this second plot in the following analyses.\\
We can now compare our observational results with the profile expected by theoretical models.\\
The model adopted to derive the expected radial $I_R-I_X$ slope was firstly introduced in \cite{Bonafede2022} and used in \cite{Balboni2024} \citep[see also][]{Santra2026}. 
It is derived in the framework of the turbulent re-acceleration scenarios for radio halo production, which has been widely studied and developed in past literature works (e.g. \citealt{Cassano2005, Brunetti-Lazarian07, Brunetti-Lazarian11}).
In particular, this model extends the formalism used to reproduce integrated quantities \citep[see, e.g.,][]{Cassano2023} to the spatially resolved case. 
It exploits the known connection between the X-ray and radio emission to derive an analytical expression for the latter, allowing the magnetic field and the CRe content to vary radially.\\
According to this modelisation, the radio emissivity can be expressed as:
\begin{equation}
    \epsilon_R \propto F \eta_e \frac{B^2}{B^2 + B^2_{IC}},
\end{equation}
where $B_{IC} \approx 3.25 (1+z)^2 ~ {\rm \mu G}$ is the CMB equivalent magnetic, $\eta_e$ is the re-acceleration efficiency and $F$ is the turbulent energy flux.
Considering an isotropic distribution of the electrons' momentum space ($f(p)$) and adopting the case of second-order acceleration with super-Alfvenic solenoidal turbulence \citep{Brunetti-Lazarian07,Brunetti-Lazarian16}, we can rewrite $\eta_e \approx \frac{U_{CRe}}{F} \frac{c^3_s M^3_t}{L v_A}$, where $U_{CRe}$ is the energy density of re-accelerated electrons, $M_t$ is the turbulent Mach number and $v_A$ is the Alfvén velocity.
Hence, for a constant temperature profile and $M_t$, and assuming a magnetic field profile that scales as $B \propto n_{th}^{0.5}$, the radio emissivity becomes:
\begin{equation}
    \label{eq:em_radio}
    \epsilon_R(r) = \frac{X(r)}{X(0)}\epsilon_{R,0} \left( \frac{\epsilon_X(r) }{\epsilon_{X,0}} \right )^{1/2} \frac{1+ \left ( \frac{B_{IC}}{B_0}\right )^2 } { 1 + \left (\frac{B_{IC}}{B_0} \right )^2 \left ( \frac{\epsilon_{X,0} }{\epsilon_X(r)} \right )^{1/2}},
\end{equation}
where $\epsilon_X$ is the X-ray emissivity and $X(r) = \frac{U_{CRe}}{U_{th}}$ is the ratio between the energy densities of CRe ($U_{CRe}$) and the thermal gas ($U_{th}$).
We computed $\epsilon_R$ for different values of the central magnetic field $B_0 = 2-6 \mu$G, spanning the range of values typically inferred or assumed for massive clusters \citep[e.g.][]{Stuardi2021}, and assuming a constant $X(r)$.
We derived $\epsilon_X$ by fitting a $\beta$-model to our X-ray images using the \texttt{PYPROFFIT} package presented in \cite{Eckert2020}. 
We finally projected both radio and X-ray emissivities obtaining the model surface brightness profiles and derived $k(r)$ using Eq.~\ref{eq:k}. 
We stress that these models are used as reference curves for comparison rather than as a fitting grid, since given the sample size and the cluster-to-cluster diversity, $B_0$ itself is expected to vary among clusters \cite[e.g.][]{Loi2026}.\\
Fig. ~\ref{fig:stacked_prof+model} (bottom panel) shows how the observed increasing $k(r)$ trend of dynamically disturbed clusters, as a function of the radius, is consistent with the model in the $\sim 0.2-0.8 \rm ~ R_{500}$ range.
The only two points that disagree with the model expectation are in the very central regions ($<0.2 ~ \rm R_{500}$) and towards the clusters outskirts ($k\sim 1.5$ at $r\gtrsim0.9 ~\rm R_{500}$; not shown here to facilitate the readability of the plot, but visible in the global profile in the top panel).
The former discrepancy is likely due to the fact that cluster cores are more affected by non-gravitational processes. 
The latter is (i) dominated by the variance due to the low number of measurements available at such distances; (ii) as in the inner parts, the model assumptions may no longer hold effectively in those regions, e.g. a constant temperature profile \citep[see][]{Rossetti2024}.\\
This comparison tells us that for more perturbed systems the averaged trend is in line with expectations in the turbulent re-acceleration theory for $B_0 \simeq 4-6 ~\mu$G, and that such a framework is not excluded by the data. In particular, a constant $U_{CRe}/U_{th}$ ratio remains a viable description of the averaged disturbed cluster profile within this framework, since a varying $X(r)$ would produce significantly different results, as discussed later in this section and in \cite{Balboni2024}.
We caution, however, that this comparison should be interpreted with care, given the number of assumptions entering the model and the limited sample size.
On the other hand, mixed systems display a markedly different behaviour, with $k(r)$ decreasing radially. 
This trend is consistent with the results of our pilot study \citep{Balboni2024}, where we showed that,  within the considered framework, it could be reproduced by introducing a radially increasing $X(r)$ profile, i.e. a radially varying $\frac{U_{CRe}}{U_{th}}$.
The sample analysed here provides additional support for this interpretation. 
This suggests that, in less dynamically disturbed systems, the spatial correspondence between the thermal and non-thermal components weakens toward larger radii, with the synchrotron-emitting component becoming progressively more extended relative to the thermal ICM.\\
This result may be also linked to the different dynamical state of these systems, resulting in different ICM properties. In fact, in mixed systems, especially toward the inner and meso-scale regions, the non-thermal plasma may be more affected by the core cycle: intermittent AGN injection, fossil plasma motions, turbulent diffusion, etc \citep[e.g.][]{Gaspari2018, Gaspari2020, Wittor2023, Barbani2026}. Within the adopted framework, this behaviour is consistent with an increasing $X(r)$, although other factors could also play a role (e.g. seed-CRe transport, fossil plasma, varying re-acceleration efficiency).\\
Finally, we remind the reader that in \cite{Balboni2024} the radial $X(r)$ trend required to match the observed $k(r)$ values was not consistent with the radio surface brightness profiles, as it was producing a brighter radio emission than the observed one. 
By finding a similar result for the mixed sub-sample, we are facing the same situation here. Therefore, we stress that an increment of the $X(r)$ alone is not enough to reproduce the whole observables, although it is required by the theoretical framework tested here. 
A possible way to weaken the requirement on $X(r)$ can be the inclusion of weaker coupling between the magnetic field and the thermal ICM.
In fact, a shallower relation, i.e. $B\propto n_{th}^{\delta}$ with $\delta<0.5$, would partially account for an increment of $X(r)$, producing a flatter radio profile. 
A more detailed treatment of the underlying relations between the thermal and non-thermal components of the ICM, and their spatial distributions, would require dedicated numerical simulations. Such an analysis is beyond the scope of this work, but could be pursued in future studies with larger samples.\\
\subsection{Radio profile scaling: theoretical expectations}\label{sec:emR_rescaling}

\begin{figure}
    \centering
    \includegraphics[width=\linewidth]{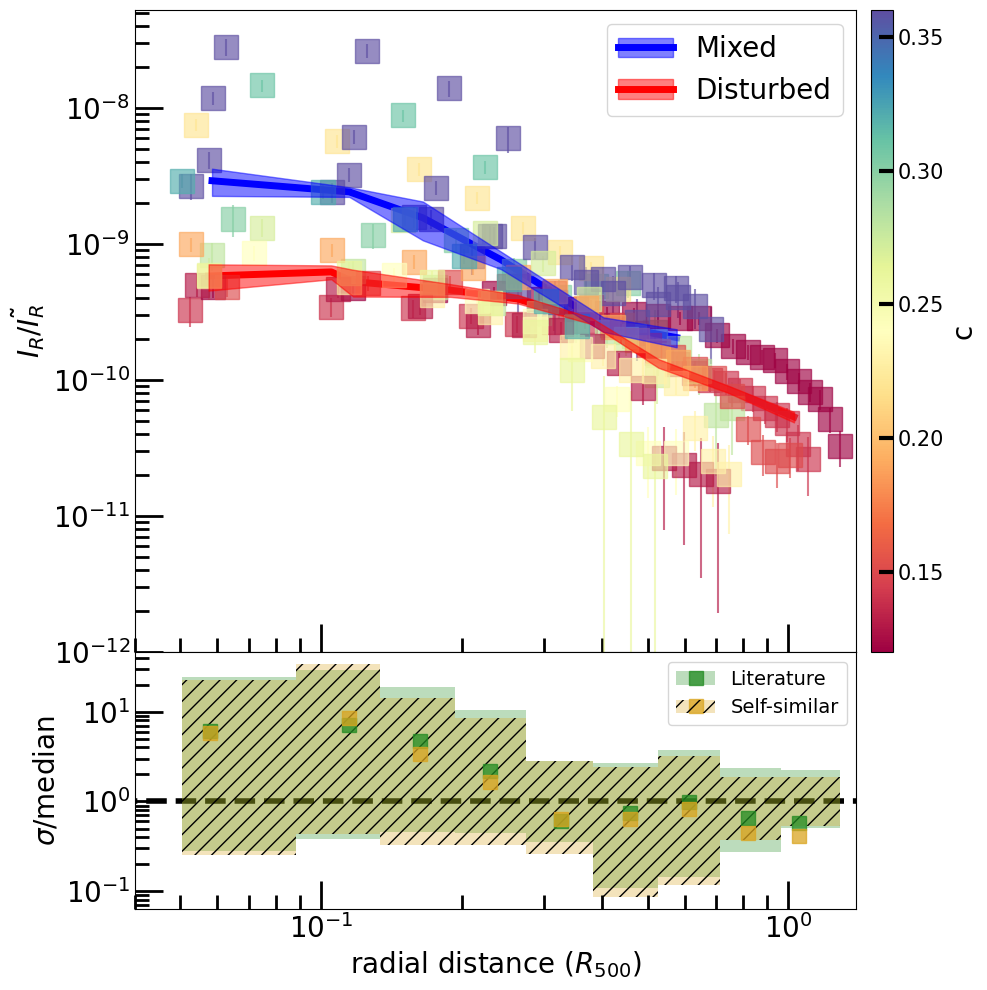}
    \caption{Radio halo profile rescaling results. Top: Scaled radio profiles of the 18 targets studied, assuming $\alpha_M=0.46$ and $\alpha_z=4.22$ and colour-coded by the light-concentration parameters $c$. Highlighted in red and blue the median profiles of mixed and disturbed targets \citep{Campitiello2022}, respectively. Bottom: Standard deviation of the rescaled profile assuming self-similar scaling (orange) and the one adopted in the top panel (green). The height of the shaded area is the range of values spanned in each bin.}
    \label{fig:scaledRprof}
\end{figure}

The model expressed in Eq.~\ref{eq:em_radio} provides a direct link between the radio and the X-ray emissivity.  Therefore, one can investigate the validity and limits of such a relation by directly comparing model predictions and observations. 
In particular, we aim to follow up the analysis we made in \cite{Balboni2025} where we obtained a consistent view of the mass and redshift scaling of the radio emission for both integrated (luminosity) and resolved (radial profiles) halo properties.
Here, we want to test whether the known scaling of $\epsilon_X$ with mass and redshift provides a valuable description of the dependence of $\epsilon_R$ (and then of the radio surface brightness $I_R$) by the same quantities.
To do so, from Eq.~\ref{eq:em_radio} we produced, for each cluster, the expected $\epsilon_R$ in which we included a mass and redshift contribute due to $\epsilon_X$.
Specifically, in Eq.~\ref{eq:k} we replaced:
\begin{equation}
    \widetilde{\epsilon_{X}} = \frac{\epsilon_X(r) }{\epsilon_{X,0}}
\end{equation}
with $\xi(M_{500},z) \times \widetilde{\epsilon_{X}}$, where
\begin{equation}
    \xi(M_{500},z) = \left(\frac{ M_{500}}{10^{14}  M_{\odot}}\right)^{\alpha_M} E_z^{\alpha_z} 
\end{equation}
so that the normalisation of the X-ray emissivity term will be modified.
Then, we projected the derived radio emissivity to obtain a pseudo-radio surface brightness $\tilde{I_R}(M_{500}, E_z)$ which we used to rescale the observed radio halo surface brightness profile $I_R$\footnote{These profiles have been obtained as detailed in \cite{Balboni2025}}.\\
To define $\alpha_M$ and $\alpha_z$, we firstly expressed the X-ray emissivity as $\epsilon_X \propto n_g^2$, with $n_g$ the gas number density (assuming the cooling function independent on the temperature in the 0.7-1.2 keV range, \citealt{Ettori2000}). 
Then, we parameterised the mass and redshift scaling of $n_g$ by considering two different cases: (i) the best-fit results found by \cite{Pratt2022} and \cite{Ettori2023}, who constrained the mass and redshift scaling of ICM thermodynamic quantities using X-ray observations, and (ii) the expectations from the self-similar scenario \citep[see, e.g.,][]{Ettori2023}.
Considering the literature result (self-similar) scaling of $n_g \propto M^{0.23} E_z^{2.11}$ ($\propto E_z^2$) we rewrite $\epsilon_X \propto M^{0.46} E_z^{4.22}$ ($\propto E_z^{4}$).\\
In the top panel of Fig.~\ref{fig:scaledRprof} we present the rescaled radio profiles ($I_R/\tilde{I_R}$) assuming the scaling relations from the literature. In the bottom panel, we report the scatter of the rescaled profiles using both methods.
\\
Similarly to what is observed for the thermal ICM \citep[e.g.][]{Pratt2009,Bartalucci23}, there is a clear, radial trend of the $I_R/\tilde{I_R}$ dispersion. From both the upper and lower panel of Fig.~\ref{fig:scaledRprof}, it is possible to identify a decrement of the scaled profile scatter moving to larger radii. 
In particular, the ratio between the standard deviation and the median value of $I_R/\tilde{I_R}$ shows a clear decrement ($\sim$ one order of magnitude) between the central and outer regions.
Furthermore, we can get clues about the origin of such scatter by colour-coding each scaled emission profile by $c$, which provides information on the amount of X-ray emission in the cluster core.
Here we see that the profiles that show a higher (more peaked) $I_R/\tilde{I_R}$ are the ones with larger $c$ in X-rays. On contrary, objects with lower $c$ present a flatter $I_R/\tilde{I_R}$ profiles in the central regions. 
\ {We also checked whether the observed features were due to the fact that higher mass/redshift clusters were also the most disturbed ones. We found no significant correlations among $\rm M_{500}$/z and $c$.}\\
This finding is consistent with the results discussed in Sec.~\ref{sec:disc_ptp}, where we showed that the clusters present a tight correlation between their thermal and non-thermal components. It is then natural to find such a correlation also in the profile features as shown here, suggesting, even more, that these two plasma components are shaped, to the first order, by the same physical mechanisms.\\
Furthermore, this dichotomy is also evident
if we separate our sample in mixed and disturbed clusters (see Sec.~\ref{sec:disc_radial_slopes}) and compute the median profile of the two subsamples. 
Additionally, if considering either single profiles or median ones, we see that the separation between less and more dynamically disturbed clusters becomes weaker at larger radii.\\
The results obtained suggest that, to first order, the spatial distribution of the synchrotron-emitting plasma is regulated by the same physical processes that shape the thermal ICM. The residual differences between mixed and disturbed systems indicate, however, that the cluster dynamical state leaves a measurable imprint on the radial distribution of the non-thermal component.
\section{Conclusions}\label{sec:summary}

We have presented the first homogeneous, spatially resolved radio–X-ray analysis for a sample of 18 radio halo clusters by combining LoTSS-DR2 \citep{Shimwell2022,Botteon2022} observations with the CHEX-MATE X-ray dataset \citep{Arnaud21}.
The excellent sensitivity and homogeneous quality of the radio and X-ray data allowed us to study, for the first time on a cluster sample, the spatial connection between the thermal and non-thermal plasma components in the ICM.
Our results can be summarised as follows:
\begin{itemize}

    \item We find a clear correlation between radio and X-ray surface brightness in all but one cluster, confirming the close connection between the thermal and non-thermal components of the ICM. In nearly $70\%$ of the sample the relation is sublinear, indicating that the synchrotron-emitting component display generally a flatter distribution than the thermal plasma. Within the turbulent re-acceleration framework, and under reasonable assumptions on the magnetic-field distribution, this behaviour is consistent with a shallower radial distribution of CRe than of the thermal gas.
    Hence, if such a sub-linear $I_R-I_X$ relation holds even at large radii, one might expect to find evidence of radio emission extending up to (and even beyond) the cluster outskirts, on broader scales than that probed by X-ray observations \citep[e.g.][]{Botteon2022-a2255, Cuciti22, Rajpurohit2025,Santra2026}.
    \item We also report a few cases where the $I_R-I_X$ relation is better described by a superlinear slope. By looking more carefully at these targets, we noted how the radio remarkably follows the X-rays, justifying such a tight, best-fit relation. We discussed how a possible explanation of their observed properties can arise from the presence of double non-thermal plasma component (inner and outer), as also found for other cases \citep[e.g.][]{Biava2024, vanWeeren2026}.
    \item Comparing all the obtained slopes from different objects, we found no correlation with the dynamical parameters (M and $c$), and a weak trend with the cluster mass and/or redshift which is worth investigating with larger samples. 
    \item We found that radio and X-ray centroid offsets are systematically smaller than the corresponding peak offsets, reflecting the close spatial correspondence between the large-scale thermal and non-thermal plasma distributions. In addition, the correlation between $\Delta_{cent}$ and the X-ray morphological indicators of the cluster dynamical state, suggests that it may provide a useful new tracer of cluster evolution.
    \item When inspecting the radial variation of the $I_R-I_X$ slope in our sample we found a wide variety of trends. This demonstrates that the relative spatial distribution of the thermal and non-thermal components is not constant across the cluster volume.
    \item Averaging the radial $k(r)$ profiles, we find clear differences between the disturbed and mixed clusters. The former are consistent with expectations from the turbulent re-acceleration framework for plausible magnetic field strengths. In particular, we observed that a constant ratio between the non-thermal and thermal energy densities remains a viable description for these systems, although given the assumptions entering the model and the sample size, this comparison should be interpreted with caution. Instead, for less disturbed clusters, a progressively increasing non-thermal-to-thermal energy ratio with radius is required to reproduce observational results. This suggests that the spatial coupling between the two plasma components evolves with the dynamical state of the cluster.
    \item Finally, we show that radio halo surface-brightness profiles can be rescaled using X-ray information together with expectations from turbulent re-acceleration models. The proposed scaling substantially reduces the diversity among individual systems and reveals common radial trends, while the remaining deviations appear to be linked to the cluster dynamical state. Rather than indicating a strict self-similarity of radio halos, these results suggest that their average radial properties are largely governed by the same large-scale physical processes shaping the thermal ICM, with merger history and particle acceleration introducing the observed cluster-to-cluster variations.
    \end{itemize}

Overall, our results provide the first systematic observational picture of the spatial connection between the thermal and non-thermal plasma in a representative sample of radio halo clusters. They demonstrate that the radio--X-ray relation contains valuable information on the distribution of relativistic particles and on the physical processes governing turbulence-driven particle acceleration. Future deep radio surveys with LOFAR, MeerKAT and the SKA, combined with high-quality X-ray observations, will make it possible to extend this analysis to larger samples and to place increasingly stringent constraints on models of particle acceleration in the ICM.

\bibliographystyle{aa}

\bibliography{biblio.bib} 

\begin{acknowledgements}
We would like to thank the anonymous referee for their helpful suggestions, which have improved the presentation of this manuscript.
We acknowledge the developers of the following Python packages which were used in this work: \textsc{ASTROPY} \citep{astropy:2013,astropy:2018,astropy:2022}, \textsc{MATPLOTLIB} \citep{matplotlib}, \textsc{SCIPY} \citep{SciPy-NMeth}, \textsc{NUMPY} \citep{numpy}, \textsc{CORNER}, (\citealt{cornerpy16}), \textsc{PyMC} \citep{pymc16,pymc_v4.2} and \textsc{ArviZ} \citep{arviz2026}.
AB and MB acknowledge support from the ERC CoG $\vec{B}$ELOVED, GA N.101169773. 
AI acknowledges support from the institutional project RVO:67985815 and the project 25-19512L of the Czech Science Foundation.
M.S. acknowledges financial contributions from contract INAF mainstream project 1.05.01.86.10, INAF Theory Grant 2023: Gravitational lensing detection of matter distribution at galaxy cluster boundaries and beyond (1.05.23.06.17), INAF Guest Observer Grant 2024: Towards anchoring the mass scale of galaxy clusters with galaxy kinematics (1.05.24.02.15), and contract Prin-MUR 2022 supported by Next Generation EU (n.20227RNLY3, The concordance cosmological model: stress-tests with galaxy clusters).
J.S. acknowledges financial support from NASA awards 80NSSC25K8006 and 80NSSC25K8809.
HB, FDL, and PM acknowledge the financial contribution from the contract Prin-MUR 2022 supported by Next Generation EU (M4.C2.1.1, n.20227RNLY3 \textit{The concordance cosmological model: stress-tests with galaxy clusters}) and the support by INFN through the InDark initiative.
M.G. acknowledges support from the ERC Consolidator Grant \textit{BlackHoleWeather} (101086804).
M.R., F.G., A.B., M. B., acknowledge the financial contribution from the INAF GO grant 1.05.24.02.10 Extended Radio Emission in Galaxy Clusters at deep focus with MeerKAT.

LOFAR \citep{LOFAR2013} is the Low Frequency Array designed and constructed by ASTRON. It has observing, data processing, and data storage facilities in several countries, which are owned by various parties (each with their own funding sources) and are collectively operated by the ILT foundation under a joint scientific policy. The ILT resources have benefited from the following recent major funding sources: CNRS-INSU, Observatoire de Paris and Université d'Orléans, France; BMBF, MIWF-NRW, MPG, Germany; Science Foundation Ireland (SFI), Department of Business, Enterprise and Innovation (DBEI), Ireland; NWO, The Netherlands; The Science and Technology Facilities Council, UK; Ministry of Science and Higher Education, Poland; The Istituto Nazionale di Astrofisica (INAF), Italy. This research made use of the Dutch national e-infrastructure with the support of the SURF Cooperative (e-infra 180169) and the LOFAR e-infra group. The Jülich LOFAR Long Term Archive and the GermanLOFAR network are both coordinated and operated by the Jülich Supercomputing Centre (JSC), and computing resources on the supercomputer JUWELS at JSC were provided by the Gauss Centre for Supercomputinge.V. (grant CHTB00) through the John von Neumann Institute for Computing (NIC). This research made use of the University of Hertfordshirehigh-performance computing facility and the LOFAR-UK computing facility located at the University of Hertfordshire and supported by STFC [ST/P000096/1], and of the Italian LOFAR IT computing infrastructure supported and operated by INAF, and by the Physics Department of Turin university (under an agreement with Consorzio Interuniversitario per la Fisica Spaziale) at the C3S Supercomputing Centre, Italy.
GWP acknowledges support from the Centre national d’études spatiales (CNES), France (ROR: https://ror.org/04h1h0y33) within the framework of the XMM-Newton mission.\\
MB thanks Simone Candini for the careful checks performed on labels and plots. 
\end{acknowledgements}

\begin{appendix}

\section{Targets properties}

\begin{table*}[ht]
    \centering
    \caption{Main properties of the 18 radio halo clusters of the crossmatched CHEX-MATE--LoTSS DR2 sample.}
    \label{tab:sample_18}
    \begin{tabular}{ccccccccc}
    \toprule
    Name & Alternative name & z & \thead{$M_{500}$ \\ $(10^{14}M_{\odot})$} & \thead{$R_{500}$ \\ (arcmin)} & M & $c$ & \thead{$\sigma_{RMS}$ \\ $({\rm \mu Jy ~ beam^{-1}})$} & \thead{$\Delta_{cent}$ \\ (kpc)} \\
    \midrule
    PSZ2G031.93+78.71 & A1775 & 0.072 & $2.72^{+0.23}_{-0.25}$ & 11.64 & -0.47 & $0.31^{+0.07}_{-0.07}$ & 583 & $156 \pm 8$ \\
    PSZ2G040.58+77.12 & A1800 & 0.075 & $2.57^{+0.21}_{-0.22}$ & 11.08 & 0.56 & $0.3^{+0.01}_{-0.01}$ & 604 & $124 \pm 30$ \\
    PSZ2G046.88+56.48 & A2069 & 0.115 & $5.1^{+0.21}_{-0.23}$ & 9.40 & 1.41 & $0.14^{+0.04}_{-0.04}$ & 469 & $209 \pm 15$ \\
    PSZ2G048.10+57.16 & A2061 & 0.078 & $3.54^{+0.2}_{-0.21}$ & 11.90 & 1.03 & $0.12^{+0.02}_{-0.02}$ & 499 & $183 \pm 9$ \\
    PSZ2G049.32+44.37 & A2175 & 0.097 & $3.76^{+0.26}_{-0.27}$ & 9.87 & 0.08 & $0.28^{+0.06}_{-0.07}$ & 451 & $43 \pm 23$ \\
    PSZ2G053.53+59.52 & A2034 & 0.113 & $5.21^{+0.2}_{-0.21}$ & 9.58 & 0.98 & $0.22^{+0.03}_{-0.04}$ & 339 & $37 \pm 3$ \\
    PSZ2G055.59+31.85 & A2261 & 0.224 & $7.72^{+0.31}_{-0.29}$ & 5.99 & -0.46 & $0.46^{+0.05}_{-0.05}$ & 270 & $11 \pm 8$ \\
    PSZ2G056.77+36.32 & A2244 & 0.095 & $4.34^{+0.19}_{-0.21}$ & 10.53 & -0.81 & $0.49^{+0.06}_{-0.06}$ & 481 & $76 \pm 27$ \\
    PSZ2G066.41+27.03 & / & 0.575 & $7.69^{+0.52}_{-0.53}$ & 2.88 & 0.74 & $0.15^{+0.01}_{-0.02}$ & 270 & $52 \pm 3$ \\
    PSZ2G077.90-26.63 & A2409 & 0.147 & $4.99^{+0.26}_{-0.25}$ & 7.46 & -1.03 & $0.4^{+0.06}_{-0.07}$ & 647 & $18 \pm 24$ \\
    PSZ2G083.29-31.03 & / & 0.412 & $7.64^{+0.4}_{-0.41}$ & 3.66 & 0.36 & $0.27^{+0.03}_{-0.03}$ & 311 & $76 \pm 4$ \\
    PSZ2G107.10+65.32 & A1758-NW & 0.280 & $7.8^{+0.26}_{-0.27}$ & 5.00 & 0.66 & $0.19^{+0.03}_{-0.04}$ & 200 & $50 \pm 4$ \\
    PSZ2G111.75+70.37 & A1697 & 0.183 & $4.34^{+0.32}_{-0.33}$ & 5.88 & 1.47 & $0.13^{+0.03}_{-0.03}$ & 211 & $99 \pm 10$ \\
    PSZ2G113.91-37.01 & / & 0.371 & $7.58^{+0.54}_{-0.56}$ & 3.96 & 0.74 & $0.23^{+0.03}_{-0.04}$ & 263 & $68 \pm 4$ \\
    PSZ2G143.26+65.24 & A1430 & 0.363 & $7.26^{+0.39}_{-0.42}$ & 3.97 & 0.87 & $0.24^{+0.03}_{-0.03}$ & 182 & $19 \pm 5$ \\
    PSZ2G179.09+60.12 & A1068 & 0.137 & $3.84^{+0.31}_{-0.34}$ & 7.26 & -1.09 & $0.65^{+0.06}_{-0.07}$ & 335 & $95 \pm 20$ \\
    PSZ2G186.37+37.26 & A697 & 0.282 & $11.0^{+0.37}_{-0.37}$ & 5.57 & -0.20 & $0.32^{+0.03}_{-0.04}$ & 301 & $7 \pm 6$ \\
    PSZ2G192.18+56.12 & A961 & 0.124 & $3.62^{+0.28}_{-0.31}$ & 7.80 & 0.19 & $0.26^{+0.07}_{-0.08}$ & 297 & $152 \pm 31$ \\
    \bottomrule
    \end{tabular}
\end{table*}

\section{$I_R - I_X$ relation}

The same radio -- X-ray surface brightness relations discussed in Sec.~\ref{sec:disc_ptp} and shown in Fig.~\ref{fig:Ix-Ir_tot} with different colour-coding.

\begin{figure}[ht]
    \centering
    \includegraphics[width=\linewidth]{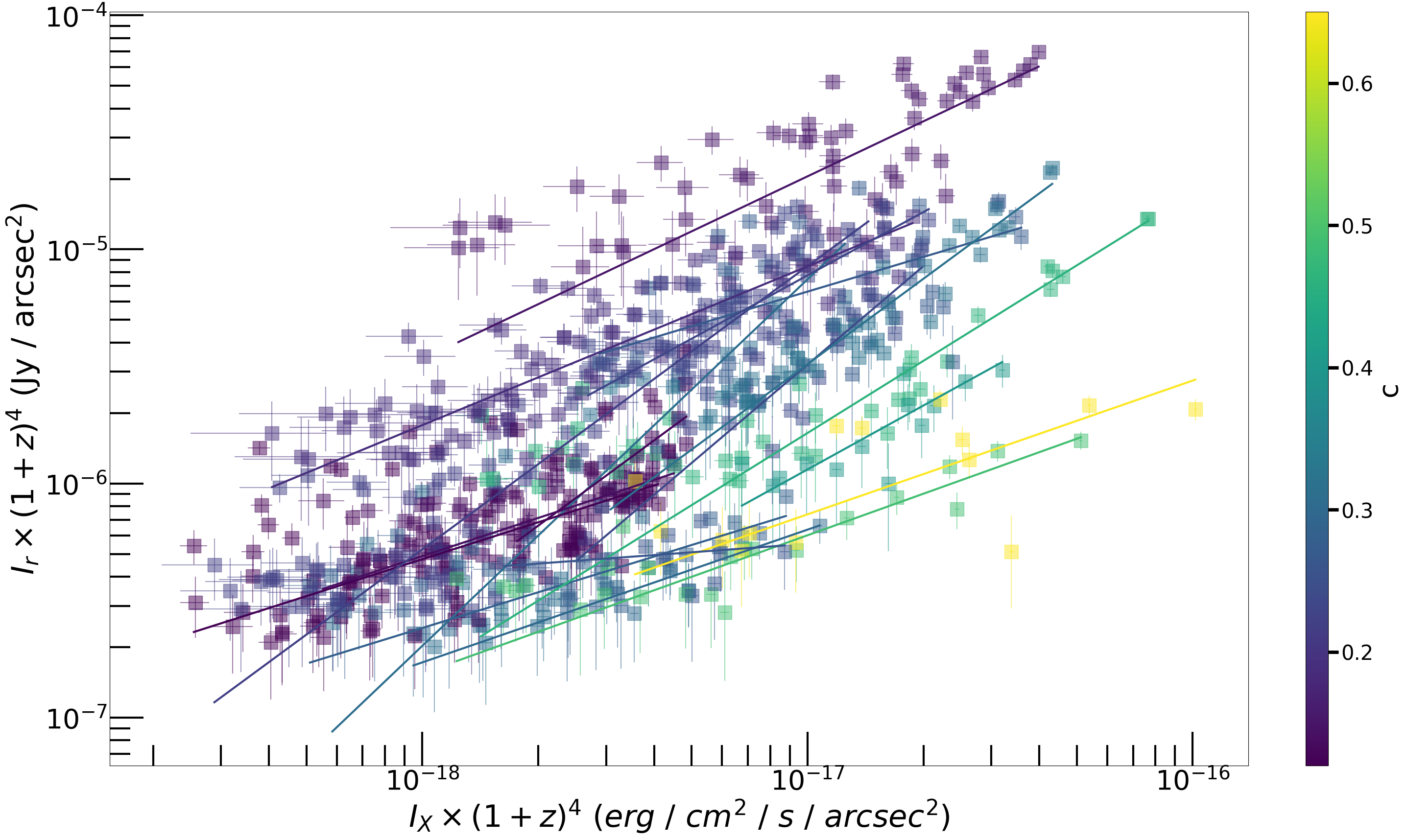}
    \caption{$I_R - I_X$ relations of all the clusters, color-coded by the light concentration $c$. Solid lines show the best-fit relations.}
    \label{fig:Ix-Ir_tot_cpar}
\end{figure}

\section{Tests on the $I_R - I_X$ analysis}

In this section we present some additional tests to asses the robustness and limits of the presented analyses on the $I_R-I_X$ relations.

\subsection{Point-to-point analysis}\label{appendix:test_ptp}

\ {Following the diagnostic tests performed firstly by \cite{Botteon2020-A2255} and subsequently by \cite{Balboni2024}, we investigated how our results can change depending on the adopted thresholds on $\sigma_{RMS}$ for selecting the radio halo regions. 
\begin{figure}[ht]
    \centering
    \includegraphics[width=\linewidth]{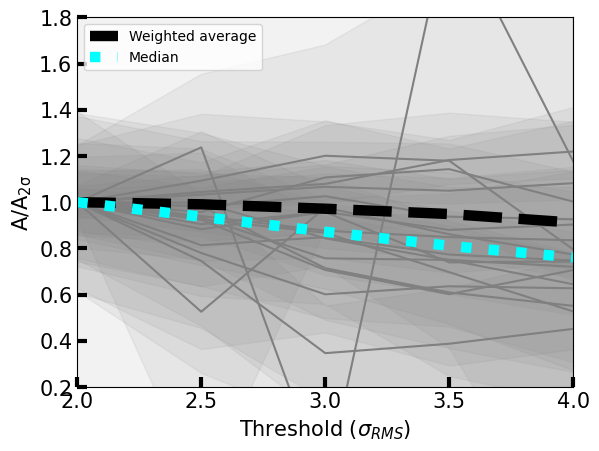}
    \caption{Relative variation of A, in units of the value obtained cutting $I_R$ at $2\sigma_{RMS}$ (see Sec.\ref{sec:ptp}), as a function of the radio threshold imposed. Each line is a different cluster, the shaded areas are the error of each measurement and black and cyan lines identify the global weighted average and median, respectively.}
    \label{fig:slope-thresh}
\end{figure}
In Fig.~\ref{fig:slope-thresh}, we report the variation of A as a function of the threshold adopted on $I_R$. As also pointed out in past works, the effect of increasing the threshold is a reduction of the slope. 
Here we see that, despite accounting for the Malmquist bias during the fitting process, when excluding the data below $>3\sigma_{RMS}$, the recovered A value can differ from the initial one by $\gtrsim 10\%$. 
This is particularly evident when considering the median value of A, while when computing the weighted average, with weights the inverse of the slope error squared ($\rm 1/(\Delta A)^2)$, the best-fit slopes agree also for different thresholds.
A further diagnostic is provided by how the relative error changes as a function of the threshold set.
\begin{figure}
    \centering
    \includegraphics[width=\linewidth]{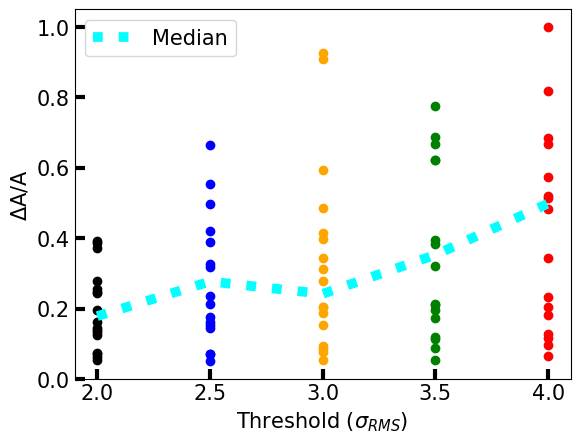}
    \caption{Relative error of A as a function of the radio threshold imposed. Each point is a different target, while the cyan line is the median trend.}
    \label{fig:relerr_thresh}
\end{figure}
We see that due to the reduction of available measurements, the error on the best-fit slope can become two times larger that the initial case. This indicates that, despite the variation of A depends on the threshold chosen, such a value become less and less reliable due to the lower statistics available (which is exactly what the difference between the median and weighted average values is telling us).\\
We finally remark that finding an increasingly sublinear slope changing the threshold for the radio emission does not affect our result of Sec.~\ref{sec:disc_ptp}. 
In fact, the main result found is that the majority of the sample show sublinear values of A. Thus, if we would select a higher threshold, the amount of objects with sublinear slopes will increase. 
In this respect, the reported fraction of $67\pm5\%$ can be seen as a lower limit.} 

\subsection{Radial $I_R - I_X$ relation}\label{appendix:test_binning}

In this section we report the same analysis as in Sec.~\ref{sec:disc_radial_slopes} about the average $k(r)$ profile but computing it using different radial binnings and testing it against systematics due to the sample size.
\begin{figure}
    \centering
    \includegraphics[width=\linewidth]{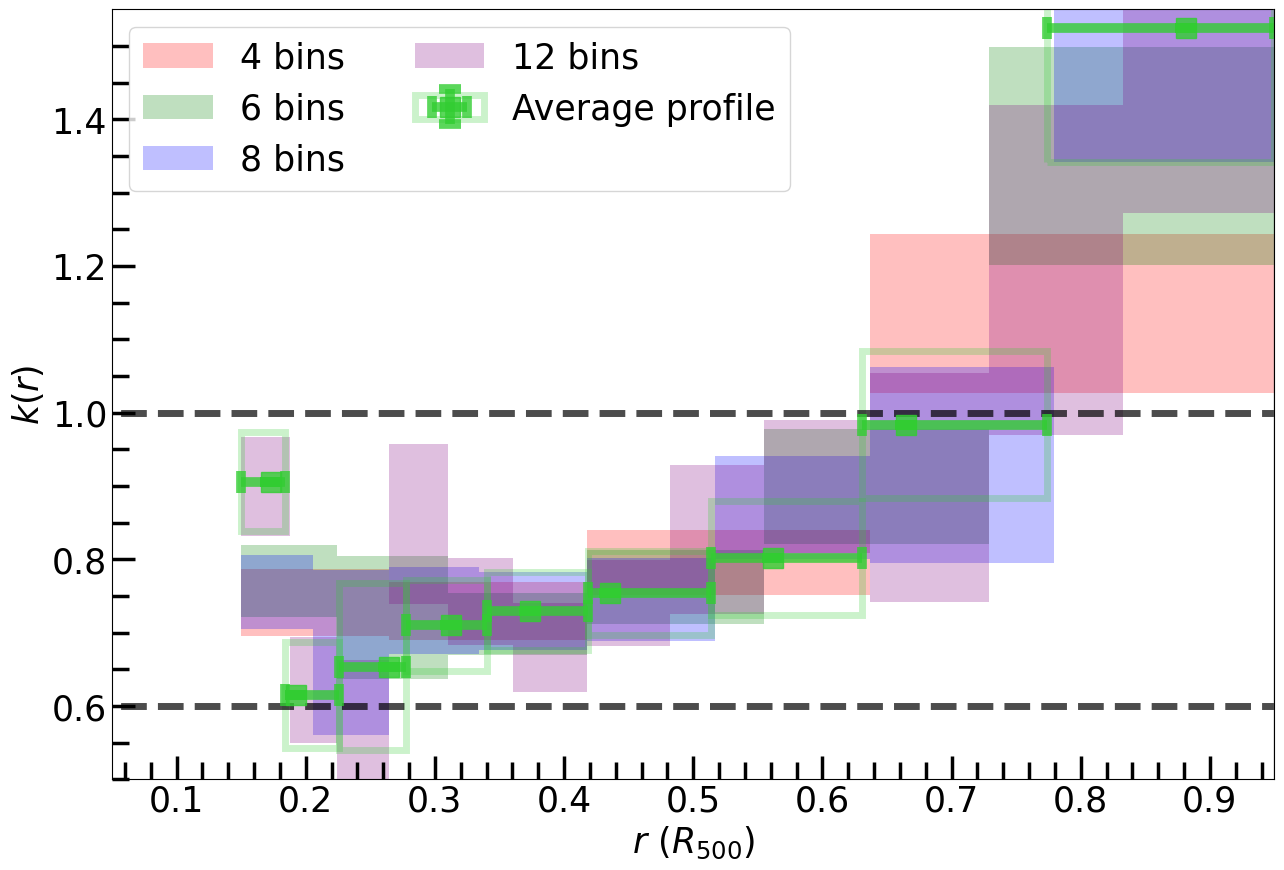}
    \caption{Weighted average profile of the $I_R - I_X$ relation slope ($k$) for our sample using different number of radial bins. The dashed lines and the average profile} are the same of Fig.~\ref{fig:stacked_prof+model} (top) for reference.
    \label{fig:stackedslope_multibin}
\end{figure}
The different trends agree with the one reported in Fig.~\ref{fig:stacked_prof+model}, showing a global increasing behaviour which becomes increasingly linear at large radii, until reaching, in the last 1-2 bins, superlinear values. The main differences occur in the central regions, where finer (broader) bins allow to resolve (to average out) the existing scatter present in the cluster centres (as found also in Sec.~\ref{sec:disc_radial_slopes}).\\
\\
To assess whether the average $k(r)$ profile presented in Fig. 8 is dominated by individual objects, we performed a leave-one-out validation test on our data. Specifically, we re-derived the average $k(r)$ profile using all targets but one and repeated the process iteratively over all the sample. The result is shown in Fig.~\ref{fig:loo_test} together with the average profile derived in Sec.~\ref{sec:disc_radial_slopes}.
\begin{figure}
    \centering
    \includegraphics[width=\linewidth]{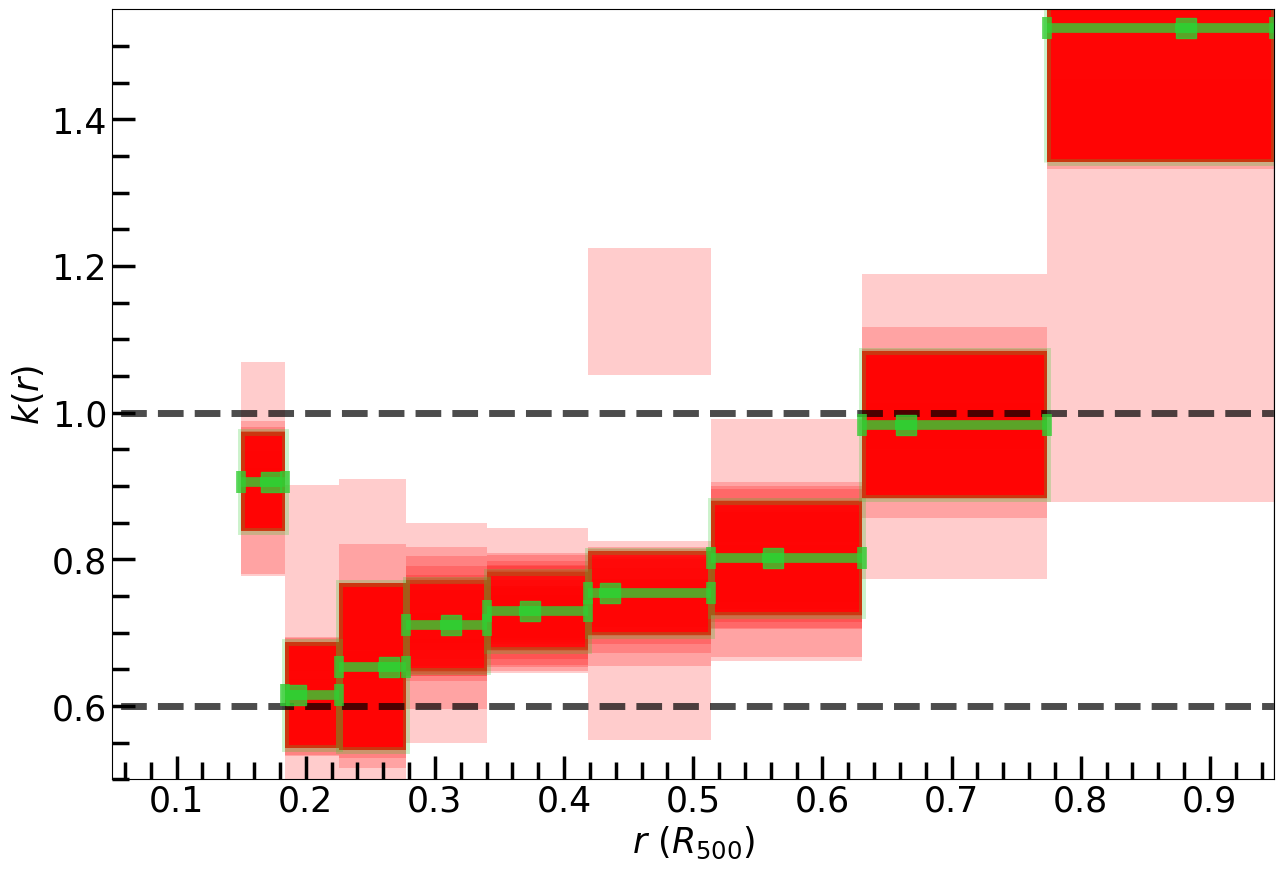}
    \caption{Weighted average profile of the $I_R - I_X$ relation slope ($k$) computed excluding iteratively one target from the whole sample (red shaded areas), together with the global profile discussed in Sec.~\ref{sec:disc_radial_slopes} (green empty blocks). The dashed lines are the same of Fig.~\ref{fig:stacked_prof+model} (top) for reference.}
    \label{fig:loo_test}
\end{figure}
Over the radial range $\sim \rm 0.15-0.8 ~ R_{500}$, the $k(r)$ replicates lie typically within, or very close to, the statistical uncertainty of the original weighted mean in every bin.
This indicates that the increasing trend observed in this range is not driven by any single cluster.
Only in the outermost bin ($\rm r \gtrsim 0.8 ~ R_{500}$), the resampled envelope spans a much wider range than the original uncertainty, reflecting the low number of measurements available at such radii.
\end{appendix}

\end{document}